%% file: ms.tex
\documentclass[usenatbib,twocolumn]{mn2e}
\usepackage[]{graphicx}
\include{aas_abbrev}

\usepackage{amssymb}

\title[Investigating time dilation in GRB light curves]{Investigating signatures of cosmological time dilation in duration measures of prompt gamma-ray burst light curves}
\author[O.M. Littlejohns et al.]
{O~.M.~Littlejohns$^{1}$,\thanks{E-mail: owenlittlejohns@gmail.com (OML)}
N.~R.~Butler$^{1}$\\
$^{1}$ School of Earth \& Space Exploration, Arizona State University,
    AZ 85287, USA\\
}

\begin{document}
\date{\today}
\pagerange{\pageref{firstpage}--\pageref{lastpage}} \pubyear{2014}
\maketitle
\label{firstpage}

\begin{abstract}
We study  the evolution with  redshift of three measures  of gamma-ray
burst (GRB) duration ($T_{\rm 90}$, $T_{\rm 50}$ and $T_{\rm R45}$) in
a fixed rest frame energy  band for a sample of 232 \textit{Swift}/BAT
detected GRBs. Binning the data  in redshift we demonstrate a trend of
increasing duration with increasing redshift that can be modelled with
a  power-law  for  all  three measures.   Comparing  redshift  defined
subsets of rest-frame duration reveals that the observed distributions
of  these  durations are  broadly  consistent  with cosmological  time
dilation.  To ascertain  if this is an instrumental  effect, a similar
analysis of \textit{Fermi}/GBM data for the 57 bursts detected by both
instruments  is  conducted,  but  inconclusive  due  to  small  number
statistics.   We  then  investigate  under-populated  regions  of  the
duration  redshift  parameter space.   We  propose  that  the lack  of
low-redshift,  long duration  GRBs is  a  physical effect  due to  the
sample being volume  limited at such redshifts. However,  we also find
that  the  high-redshift, short  duration  region  of parameter  space
suffers  from   censorship  as   any  \textit{Swift}  GRB   sample  is
fundamentally defined  by trigger criteria determined  in the observer
frame energy band of \textit{Swift}/BAT. As a result, we find that the
significance  of any evidence  for cosmological  time dilation  in our
sample of duration measures typically reduces to $<2\sigma$.
\end{abstract}


\section{Introduction}
\label{sec:intro}

Gamma-ray   burst  (GRB)   prompt  emission   has  been   detected  at
high-energies           for           over          40           years
\citep{2014ApJS..211...13V,2011ApJS..195....2S,2006ApJS..166..298K,2009ApJS..180..192F,1973ApJ...182L..85K}. It
is only  in the last decade,  however, that a  significant fraction of
detected  GRBs  have sufficient  ground-based  follow-up  to obtain  a
redshift  measurement. This  is largely  thanks to  the \textit{Swift}
satellite \citep{2004ApJ...611.1005G}, which combines the capabilities
of     its    wide     field    Burst     Alert     Telescope    (BAT;
\citealt{2005SSRv..120..143B}) with arcsecond positional accuracies of
the X-Ray Telescope  (XRT; \citealt{2005SSRv..120..165B}).  With these
X-ray  positions, ground-based facilities  have been  able to  build a
comprehensive  sample  of GRBs  with  associated  redshift using  both
photometric  and  spectroscopy methods  in  the  optical  and near  IR
wavelength regimes (e.g. \citealt{2012ApJ...756..187H}).\par

Knowing the  redshift associated with a GRB  places strong constraints
on many  properties of the transient  event. Indeed, it  was the first
GRB  redshift that finally  settled the  debate regarding  whether the
transients    were     Galactic    or    cosmological     in    origin
\citep{1997Natur.387..878M}.    Even   with  ground-based   telescopes
dedicated to GRB follow-up, and target of opportunity (ToO) programmes
in  place on large  aperture facilities,  approximately two  thirds of
\textit{Swift} GRBs do not  have an associated redshift. Additionally,
other  high-energy instruments  such  as the  Gamma-ray Burst  Monitor
(GBM;  \citealt{2009ApJ...702..791M}) on the  \textit{Fermi} satellite
cannot  provide  burst locations  with  sufficient  accuracy to  allow
narrow field ground-based facilities to obtain a redshift.\par

With such a  large fraction of GRBs lacking  redshift, searches within
the high-energy prompt light curves for tracers of redshifts have been
previously  attempted. As  GRBs  occur at  cosmological distances  and
share a common  central engine, it might expected  that a signature of
cosmological time dilation would  be measurable in these light curves.
Previous studies have considered  variability of the high-energy light
curve \citep{2001ApJ...552...57R}  and the  time lag between  the same
morphological light curve structure being observed in different energy
bands \citep{2000ApJ...534..248N}  as an indicator  of intrinsic burst
luminosity.\par

More recently, \citet{2013ApJ...778L..11Z} have considered traditional
measures    of    duration,   $T_{\rm    90}$    and   $T_{\rm    50}$
\citep{1993ApJ...413L.101K}, of a sample of \textit{Swift}/BAT GRBs in
a fixed rest frame energy band.  $T_{\rm 90}$ and $T_{\rm 50}$ are the
intervals over which  the central 90\% and 50\%  of prompt fluence are
accumulated, respectively.  This  approach differs from most attempted
duration correlations, as other  time dilation searches often consider
a fixed  energy band in  the observer frame.  Using  cross correlation
function (CCF) analyses,  it has long been known  that the typical GRB
light  curve  evolves such  that  it  becomes  softer at  later  times
\citep{2012MNRAS.419..614U,2010MNRAS.406.2149M,2002ApJ...579..386N,2000ApJ...534..248N}. Measuring
durations such  as $T_{\rm  90}$ of the  same GRB in  different energy
bands, with characteristic energy  $E$, therefore produces a different
value  of   that  duration,   where  $T_{\rm  90}   \propto  E^{-0.4}$
\citep{1996ApJ...459..393N}.\par

Measuring durations of  a sample of GRBs in  an observer frame defined
energy band  therefore combines two redshift  dependent effects. First
is  that of  cosmological  time dilation,  which  causes durations  to
increase  by  a  factor  of   $\left(  1  +  z  \right)$  as  redshift
increases. Superimposed  upon this  is also the  effect of  sampling a
different region  of the rest frame  spectrum of each  GRB. Indeed, as
$T_{\rm 90} \propto E^{-0.4}$, it is to be expected that an additional
factor of $\left(  1 + z \right)^{-0.4}$ would  affect any correlation
with redshift,  as this is required  to ensure the same  region of all
rest frame spectra are being  sampled.  Thus, by measuring duration in
an  energy  band  that  is   fixed  in  the  observer  frame,  as  has
traditionally   been   attempted,   it   is  expected   that   $T_{\rm
  90}\left(E_{\rm  1,obs}-E_{\rm 2,obs}\right)  \propto \left(  1  + z
\right)^{0.6}$.  It is  perhaps a combination of this  weakness in the
correlation strength  of duration increasing  with increasing redshift
and  the   large  intrinsic  scatter   in  the  GRB   prompt  duration
distribution that has  prevented a clear detection of  a time dilation
signature in observer frame properties.\par

By   choosing   an   energy   band   defined  in   the   rest   frame,
\citet{2013ApJ...778L..11Z} remove  the energy dependent  effects, and
thus sample  the same part  of the rest  frame spectra of all  GRBs in
their sample.   Using a rest  frame defined energy band,  the expected
correlation  only depends  on  cosmological time  dilation, such  that
$T_{\rm    90}\left(E_{\rm1,rest}/(1+z)-E_{\rm    2,rest}/(1+z)\right)
\propto \left( 1  + z \right)$. With a  standardised rest-frame energy
band  in hand, \citet{2013ApJ...778L..11Z}  then average  across broad
bins in redshift  and study the evolution of  this average $T_{\rm 90,
  mean}$.   For a  sample of  139 \textit{Swift}/BAT  GRBs,  they find
$T_{\rm  90,  mean} =  10.5\left(  1  +  z_{\rm mean}  \right)  ^{0.94
  \pm0.26}$ with  a Pearson correlation coefficient  of $r~=~0.93$ and
chance probability of $p = 7  \times 10^{-3}$.  An index of this value
is  remarkably   close  to   that  expected  from   cosmological  time
dilation.\par

In previous work,  \citet{2013MNRAS.436.3640L} examined simulations of
real  \textit{Swift}/BAT GRBs  placed at  redshifts higher  than those
they were observed at. This  work showed that the measured duration of
an  individual  GRB  evolved  with  simulated redshift  due  to  three
effects. The first  was the expected time dilation  of features within
the light curve. In addition to this, however, was the gradual loss of
the final  FRED pulse  tail due to  poorer signal-to-noise  ratio, and
eventually the  complete loss  of late time  pulses.  As such,  if the
distribution  of intrinsic  GRB durations  were constant  in  the rest
frame,  the  observed evolution  of  the  duration  distribution as  a
function  of  redshift  may  not   be  expected  to  follow  a  simple
power-law.\par

Additionally, by  conserving the  energy range of  the GRB  within the
source   frame,  each   burst  samples   a  different   part   of  the
\textit{Swift}/BAT        effective       area        curve.        As
\citet{2013ApJ...778L..11Z}  note,  the  effective  area  of  the  BAT
instrument reduces rapidly at  $E > 100$~keV and $E <  25$~keV. With a
standard  band  of $140/\left(  1  +  z  \right)$--$350/\left( 1  +  z
\right)$~keV, this  could affect the durations measured  for GRBs with
the highest and lowest redshifts. These GRBs play the most significant
role in  determining the  value of power-law  index fitted  to $T_{\rm
  90}$ as a function of redshift.\par

In  this work  we  aim to  investigate  the origins  of any  potential
duration correlations  with redshift.  In  \S~\ref{sec:data} we detail
the  sample of  GRBs used  in  this work  and the  algorithms used  to
calculate the durations analysed in this work. In \S~\ref{sec:anal} we
begin by comparing our results to those of \citet{2013ApJ...778L..11Z}
before  extending their  sample to  include 93  more recent  BAT light
curves. We also  attempt to verify if the  observed durations are real
and not  due to instrumental  effects, by analysing light  curves from
\textit{Fermi}/GBM.   In \S~\ref{sec:disc}  we then  discuss potential
sources for apparent relations between duration and redshift.\par

\section{Data}
\label{sec:data}

In this work we made use  of data from both the \textit{Swift}/BAT and
\textit{Fermi}/GBM.   In  addition  to   this  we   required  redshift
measurements,  which  were obtained  from  the \textit{Swift}  archive
(http://swift.gsfc.nasa.gov/archive/grb\_table/).\par

Of  the 863 \textit{Swift}/BAT  detected GRBs  that occurred  prior to
2014  April  24$^{\rm  th}$,  251  have  redshifts  available  in  the
\textit{Swift} archive.   Our final sample of GRBs  is reduced further
when   considering  only   long  GRBs   ($T_{\rm  90}\left(15-350~{\rm
  keV}\right)$  $\geqslant$  2  seconds)  bright  enough  to  yield  a
measurable    $T_{\rm   90}$   in    the   \citet{2013ApJ...778L..11Z}
$140/\left(1+z\right)$--$350/\left(1+z    \right)$~keV   band.    When
imposing  these  criteria, the  final  sample  consisted  of 232  long
GRBs. Short  GRBs are  excluded in  this study as  they derive  from a
different   progenitor  population   \citep{2007PhR...442..166N}.   By
definition,  they are also  short in  duration ($T_{\rm  90}\left(15 -
350~{\rm keV}\right) <  2$ seconds; \citealt{1993ApJ...413L.101K}) and
tend to  have low measured  redshifts due to  the more rapid  decay of
their  optical afterglows.  Thus,  the inclusion  of short  GRBs would
artificially enhance the strength of any positive trend in duration as
a function of redshift.\par

Of  the 232  long GRBs  selected, 89  also fulfil  the 1  second peak
photon  flux  criterion  $F_{\rm pk}$~$\geqslant$  2.6~photons.s$^{\rm
  -1}$.cm$^{\rm -2}$ as  introduced in \citet{2012ApJ...749...68S} and
used in \citet{2013ApJ...778L..11Z}.   This corresponds to an increase
in  sample size  of 67\%  and  41\% in  the full  and bright  samples,
respectively, when  compared to \citet{2013ApJ...778L..11Z}.  Finally,
of these  232 GRBs, there were also  \textit{Fermi}/GBM data available
for 57 GRBs.\par

\textit{Swift}/BAT data were downloaded from the UK Swift Science Data
Centre  (UKSSDC; http://www.swift.ac.uk/swift\_portal/). The  data for
each  burst  were then  processed  using  the  standard software  {\sc
  batgrbproduct}. This  produced event lists, from  which light curves
in user-defined energy ranges could be calculated.\par

To  produce \textit{Swift}/BAT light  curves in  the $140/\left(1  + z
\right)$--$350/\left(1  + z  \right)$~keV  energy range,  we used  the
standard {\sc batbinevt}  routine, which creates background subtracted
light curves  normalised by the number of  fully illuminated detectors
at all times. In all instances light curves were binned at 64~ms.\par

\textit{Fermi}/GBM data were downloaded from the online \textit{Fermi}
GRB
catalogue\footnote{http://heasarc.gsfc.nasa.gov/W3Browse/fermi/fermigbrst.html}. This
provided event lists  for each GRB in all of the  12 sodium iodide (NaI)
detectors.  To produce  64~ms light curves the detectors  in which the
GRB was brightest  had to be selected.  Typically  three NaI detectors
were used for each burst. For those bursts that occurred prior to 2012
July 11$^{\rm  th}$ we used the  detectors outlined in Table  7 of the
second  \textit{Fermi}/GBM GRB  catalogue \citep{2014ApJS..211...13V}.
For GRBs after  this date that we inspected  the GBM Trigger Quick-look
Plot obtained  from the online catalogue to  determine which detectors
to use.\par

\textit{Fermi}/GBM light  curves were  produced using the  event lists
from the  detectors in which the  GRB was bright.  Only counts arising
from  photons  in the  $140/\left(1  +  z  \right)$--$350/\left(1 +  z
\right)$~keV band  were extracted.  These  were then summed to  form a
single 64~ms  light curve. Periods  of burst activity  were identified
using the {\sc getburstfit}  routine available from the \textit{Fermi}
Science    Support   Center\footnote{http://fermi.gsfc.nasa.gov/ssc/}.
This routine  fits pulse  shaped Bayesian blocks  to the  light curve.
Each  pulse  is   the  sum  of  two  exponentials,   as  described  in
\citet{2005ApJ...627..324N}.\par

A background was fitted to the  region of the light curve prior to the
first Bayesian block and after  the last Bayesian block. We considered
a  constant,  linear  and   quadratic  background  and  minimised  the
$\chi^{2}$ fit  statistic for all three models.   These fit statistics
were compared using  an F-test, first between the  constant and linear
fit.  If a linear term did  not provide a 3$\sigma$ improvement to the
background  fit, the  constant background  was adopted.   Otherwise, a
second  F-test  between  the  quadratic  and linear  models  was  also
performed. In this instance, if the quadratic model was found to offer
a 3$\sigma$ improvement to the  background fit, this was then adopted,
otherwise  the  linear model  was  used.   The statistically  favoured
background was then subtracted from the entire light curve.\par

\subsection{Durations}
\label{sec:method}

\citet{2013ApJ...778L..11Z} use  the Bayesian blocks  {\sc battblocks}
algorithm, supplied  as part of  the suite of  standard \textit{Swift}
software,   to  find   $T_{\rm  90}$   in  the   $140/\left(  1   +  z
\right)$--$350/\left(  1  +  z  \right)$~keV  band.  {\sc  battblocks}
requires  a  standard  light  curve  files as  produced  by  the  {\sc
  batbinevt} routine.\par

In    this   work,    we    use   the    methodology   described    in
\citet{2007ApJ...671..656B} to determine all durations ($T_{\rm 100}$,
$T_{\rm   90}$,   $T_{\rm   50}$   and   $T_{\rm   R45}$)   for   both
\textit{Swift}/BAT and \textit{Fermi}/GBM  light curves.  $T_{\rm 90}$
and $T_{\rm 50}$ measure the central 90\% and 50\% of cumulated source
counts, respectively  \citep{1993ApJ...413L.101K}, while $T_{\rm R45}$
is the total time spanned by the bins containing the brightest 45\% of
the GRB source counts \citep{2001ApJ...552...57R}. Error estimates for
$T_{\rm  90}$,  $T_{\rm  50}$  and  $T_{\rm  R45}$  were  obtained  by
performing a  bootstrap Monte Carlo  \citep{1993stp..book.....L} using
the  counts and  associated  errors  of each  light  curve bin  within
$T_{\rm 100}$.\par

\begin{figure}
  \begin{center}
    \includegraphics[width=8.5cm,height=8.2cm,clip,angle=0]{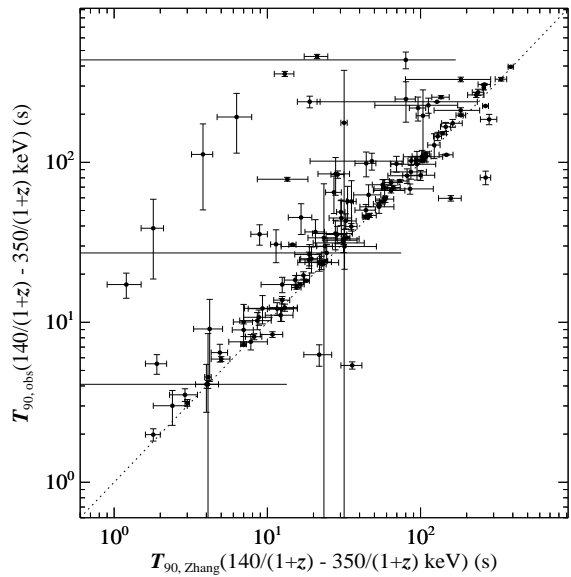}
  \end{center}
  \caption{Comparison between observer  frame $T_{\rm 90,obs}$ values
    derived   in  this   work  and   those   shown  in   Table  1   of
    \citet{2013ApJ...778L..11Z}.   For  each  GRB  light  curves  were
    extracted in the $140 / \left( 1 + z \right)$--$350 / \left( 1 +
    z \right)$~keV  energy ranges.  The dashed line  denotes equality.}
  \label{fig:obs_t90_comp}
\end{figure}

To ensure  consistency with  previous work, we  compare the  values of
$T_{\rm    90}$     found    in     this    work    to     those    of
\citet{2013ApJ...778L..11Z}.   Generally, there  is  a good  agreement
between the  two values.  For  approximately 5\% of the  population we
recover a  significantly longer  value of $T_{\rm  90}$ than  the {\sc
  battblocks}  value reported  by  \citet{2013ApJ...778L..11Z}. Visual
inspection of  these cases revealed three instances  of precursors not
detected  by the  {\sc  battblocks} algorithm,  with  the other  light
curves having a low flux  extended emission tail. The {\sc battblocks}
routine  is less  sensitive  to such  emission  tails as  they do  not
conform  to the  expected  Fast Rise  Exponential  Decay (FRED)  pulse
shape.\par

\section{Analysis}
\label{sec:anal}

\subsection{Choosing an average}
\label{sec:zhang_chk}

\begin{figure}
  \begin{center}
    \includegraphics[width=8.5cm,height=8.2cm,clip,angle=0]{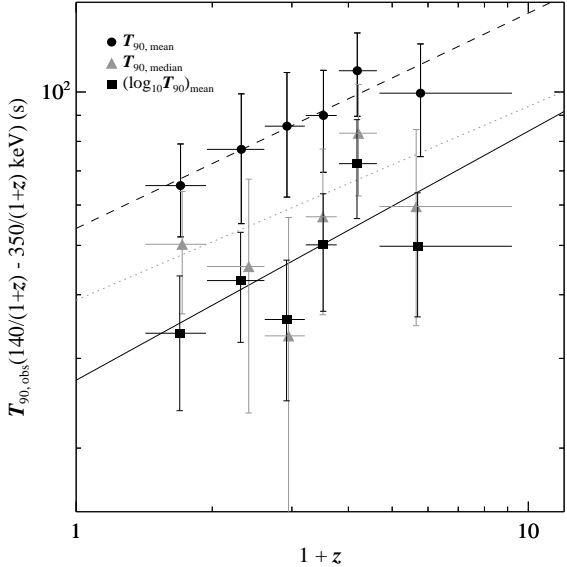}
  \end{center}
  \caption{Fitted  power-laws to observer  frame $T_{\rm  90,mean}$,
    $T_{\rm 90,median}$ and  $\left( \log_{10} T_{\rm 90}\right)_{\rm
      mean}$   for  the   full  139   GRB  \citet{2013ApJ...778L..11Z}
    sample. $T_{\rm 90,mean}$ bins  are represented by black circles,
    $T_{\rm  90,median}$ bins  are represented  by grey  triangles and
    $\left(   \log_{10}  T_{\rm   90}\right)_{\rm   mean}$  bins   are
    represented by black squares.}
  \label{fig:zhang_all}
\end{figure}

To  uncover  a  signature  of  cosmological  time  dilation  in  burst
durations,    we   first    recreated   the    models    outlined   in
\citet{2013ApJ...778L..11Z}. We began by  binning the 139 burst sample
considered     in    that     work.    In     binning     the    data,
\citet{2013ApJ...778L..11Z} measure the mean redshift and mean $T_{\rm
  90,obs}$  of the  bursts  within each  bin,  where $T_{\rm  90,obs}$
corresponds to  $T_{\rm 90}$  in the observer  frame. As the  data are
modelled with a power-law, we  consider the arithmetic mean to be more
sensitive to outliers within the bin than other averages. With this in
mind,  we also  considered the  median and  geometric mean  of $T_{\rm
  90,obs}$  within  each  bin.   The  value of  average  redshift  was
calculated  using the  same method  as $T_{\rm  90,obs}$ in  all three
instances. The  bins obtained using  these three types of  average for
the  full  \citet{2013ApJ...778L..11Z}  sample  are  shown  in  Figure
\ref{fig:zhang_all}.\par

With average bins  in hand, we then fitted power-laws  to the data, as
expected if the evolution  in the $T_{\rm 90,obs}$ distribution arises
purely from cosmological time dilation.  To assess the quality of each
fit  we  used  the  $\chi^{2}$  fit statistic,  where  the  $\chi^{2}$
minimisation was undertaken in logarithmic space.\par

In Table \ref{tab:zhang_fit_details} we detail eight alternative fits.
When modelling  average bins, we  used the statistical  scatter within
each  bin  as an  estimate  of  the error  on  each  average. For  the
arithmetic  mean and  median,  this scatter  was  calculated in  linear
space,   whilst    for   the    geometric   mean   the    scatter   in
$\log_{10}\left(T_{\rm 90,obs}\right)$ was used.\par

The first three models in Table \ref{tab:zhang_fit_details} correspond
to the  dashed lines in  Figure \ref{fig:zhang_all} as denoted  by the
key. The power-law indices  for all average methods under-predict that
expected from  cosmological time dilation,  although the error  on the
power-law index  in all three  instances is large. The  $\chi^{2}$ fit
statistic also appears to be reasonable, although inspection of Figure
\ref{fig:zhang_all}  shows  that this  is  a  consequence  of a  large
scatter  in the  $T_{\rm 90,obs}$  values within  each bin  leading to
large errors  associated with the  average bins. This  is particularly
the  case when  using the  median to  average the  values  within each
bin. Further inspection of Figure \ref{fig:zhang_all} reveals that the
median average does not conform well to the modelled power-law. \par

\begin{figure}
  \begin{center}
    \includegraphics[width=8.5cm,height=8.2cm,clip,angle=0]{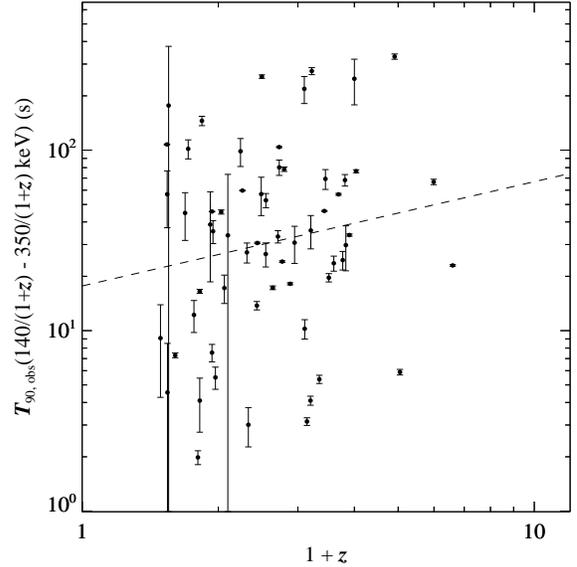}
  \end{center}
  \caption{$T_{\rm 90,obs}$ in the  $140/\left( 1 + z \right)$--$350 /
    \left( 1 +  z \right)$~keV band as a function  of redshift for the
    63 GRBs defined  in the \citet{2013ApJ...778L..11Z} bright sample.
    The dashed  line is the  \citet{2013ApJ...778L..11Z} power-law fit
    to each  individual GRB,  when all bursts  are considered  to have
    equal fractional  errors, $\left(  \Delta T_{\rm 90,obs}  / T_{\rm
      90,obs}\right)  = \left(  1 /  \ln10\right)$, not  including the
    measurement error determined by our $T_{\rm 90}$ routine.}
  \label{fig:bright_zhang_ind}
\end{figure}

We   next  recreated  the   \citet{2013ApJ...778L..11Z}  fit   to  the
individual  bright  GRBs.  This  subset  of bursts  is  identified  by
imposing   a   brightness   threshold   on  the   sample.    As   with
\citet{2013ApJ...778L..11Z}  we  apply  a  threshold  of  $F_{\rm  pk}
\geqslant        2.6$~photons.s$^{\rm        -1}$.cm$^{\rm        -2}$
\citep{2012ApJ...749...68S}, where $F_{\rm pk}$ is the one second peak
photon   flux.     This   bright    subset   is   shown    in   Figure
\ref{fig:bright_zhang_ind}.\par

We  attempted to model  the bright  \citet{2013ApJ...778L..11Z} sample
first by  considering the error in  each burst to be  that reported by
the  \citet{2007ApJ...671..656B}  $T_{\rm 90}$ algorithm. This  did  not
agree  with  the  model  fit  in  \citet{2013ApJ...778L..11Z},  so  we
repeated the  fitting, this time applying a  constant fractional error
of  $\left( \Delta  T_{\rm 90,obs}  / T_{\rm  90,obs} \right)  = 1/\ln
\left(  10  \right)$  to   every  point.   The  model  parameters  and
associated  errors obtained with  this latter  fit corresponds  to the
values reported by \citet{2013ApJ...778L..11Z}.\par

In  Figure  \ref{fig:bright_zhang_ind} we  show  the  fit obtained  by
\citet{2013ApJ...778L..11Z}.   Due to the  selected value  of constant
error for all of the data, the $\chi^{2}$ fit statistic for this model
appears reasonable. However, Figure \ref{fig:bright_zhang_ind} clearly
shows  that the  scatter in  the distribution  of $T_{\rm  90,obs}$ is
large and not accounted for by  the power-law fit.  The poor nature of
the fit is more clear when considering the fit statistic obtained when
using the true measured error in each $T_{\rm 90,obs}$.\par

As  with the  full 139  burst sample,  we also  binned the  bright GRB
sample from \citet{2013ApJ...778L..11Z}. We again fitted power-laws to
the arithmetic mean, median and geometric mean of $T_{\rm 90,obs}$ for
the bright subset of GRBs, as  shown in the bottom three rows of Table
\ref{tab:zhang_fit_details}.  The  values of power-law  index obtained
for  the  bright  subset  are  consistent, within  error,  with  those
obtained for the full sample,  although in all three cases are steeper
for  the  bright   GRB  sample.   We  note,  however,   that  $\chi  /
\nu^{2}$~$<$~1, which  indicates that  the statistical error  for each
bin is large, meaning the power-law fit is not strongly constrained in
each case.\par

\begin{table*}
  \centering
  \caption{Details of fits to $T_{\rm  90,obs}$ as a function of $1+z$
    for  the  original  bright  and  full  GRB  samples  described  in
    \citet{2013ApJ...778L..11Z}.   The  top  three rows  describe  the
    models of  the full sample  weighting each bin by  its statistical
    error on  the bin (see  Figure \ref{fig:zhang_all}). The  next two
    rows   are  for  the   bright  sample   with  each   burst  fitted
    individually. The  final three  rows are for  average bins  of the
    bright   sample,   using   statistical   error  to   weight   each
    bin. $\log_{10}N$ is the  logarithm of the fitted normalisation of
    each power-low.}
  \label{tab:zhang_fit_details}
  \begin{tabular}{cccccc}
    \hline 
    \hline
    Sample & Average &  Error & $\log_{10}N$  & Index & $\chi^{2} / \nu$  \\ 
    \hline
    All & Arithmetic  mean & $\sigma_{\rm bin}/{\rm  SE}_{\rm bin}$ 
    &  1.73 $\pm$0.05  & 0.42  $\pm$0.09 & 0.64/4 \\
    All  & Median & $\sigma_{\rm bin}/{\rm  SE}_{\rm bin}$ & 1.59 $\pm$0.12 
    & 0.38 $\pm$0.23 & 1.94/4 \\
    All & Geometric mean & $\sigma_{\rm  bin}/{\rm SE}_{\rm  bin}$  
    & 1.43  $\pm$0.13 &  0.49 $\pm$0.25 & 3.13/4 \\
    \hline
    Bright & None & $T_{\rm 90,obs} / \ln\left( 10  \right)$ & 1.25 $\pm$0.20 
    & 0.58 $\pm$0.45  & 87.76/59 \\
    Bright  & None & $\Delta  T_{\rm 90}$ & 1.89  $\pm$0.09 & -0.51 $\pm$0.21 
    & 70608.81/59  \\
    \hline
    Bright & Arithmetic  mean & $\sigma_{\rm  bin}/{\rm SE}_{\rm  bin}$  
    & 1.48  $\pm$0.10 &  0.66 $\pm$0.23  & 1.04/4 \\
    Bright &  Median &  $\sigma_{\rm bin}/{\rm SE}_{\rm  bin}$  & 1.31 $\pm$0.18  
    &  0.76  $\pm$0.43 &  1.27/4 \\
    Bright & Geometric mean & $\sigma_{\rm bin}/{\rm SE}_{\rm bin}$
    & 1.20 $\pm$0.15 & 0.72 $\pm$0.33 & 1.88/4 \\ 
    \hline
  \end{tabular}
\end{table*}

For all  further fitting in this work,  we have chosen to  only fit to
binned data.  As  we are fitting a power-law  in logarithmic space, we
adopt the geometric mean and weight all average bins by the scatter in
$\log_{10}\left(T_{\rm 90,obs}\right)$.\par

\subsection{Updating the \textit{Swift}/BAT sample}
\label{sec:zhang_ext}

The original  \citet{2013ApJ...778L..11Z} sample contains  GRBs with a
known  redshift detected  by 2012  March. We  extended this  sample by
considering all bursts  detected up to 2014 April  23$^{\rm rd}$. This
gave  us an  initial list  of  251 \textit{Swift}  detected GRBs  with
redshift.  Using our $T_{\rm 90}$ algorithm we found only 238 GRBs had
a                          measurable                          $T_{\rm
  90,obs}\left(140/\left(1+z\right)-350/\left(1+z\right)~{\rm
  keV}\right)$.   Additionally, 6  of these  GRBs were  short ($T_{\rm
  90,obs}\left( 15-350~{\rm  keV}\right) <  2$~s), so we  removed them
from  the  extended  sample.   Imposing the  $F_{\rm  pk}$  brightness
threshold \citep{2012ApJ...749...68S} on this increased subset yielded
an extended bright sample of 89 GRBs.\par

\begin{figure}
  \begin{center}
    \includegraphics[width=7.5cm,height=7.2cm,angle=0,clip]{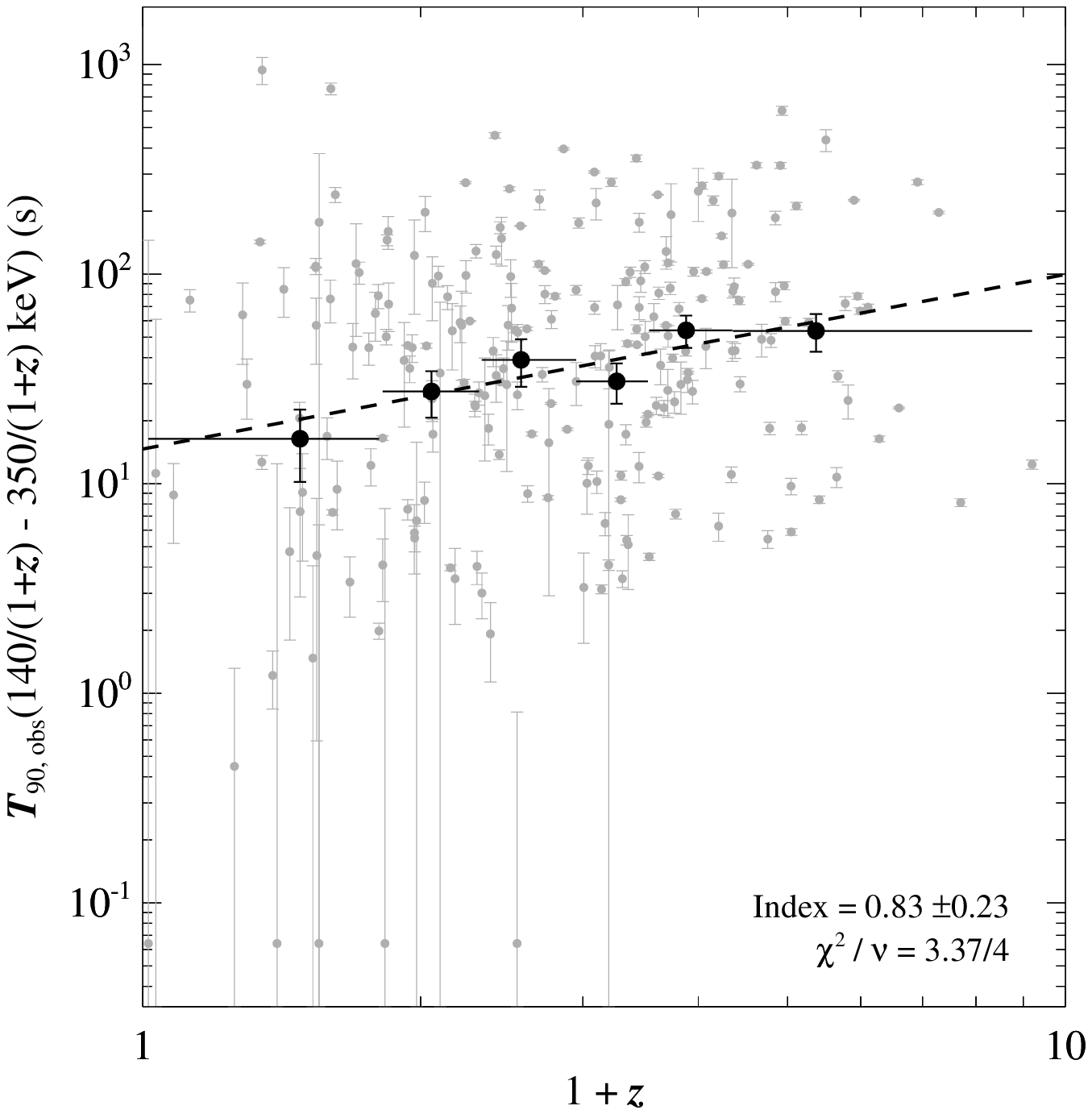}
    \\[-2ex]
    \includegraphics[width=7.5cm,height=7.2cm,angle=0,clip]{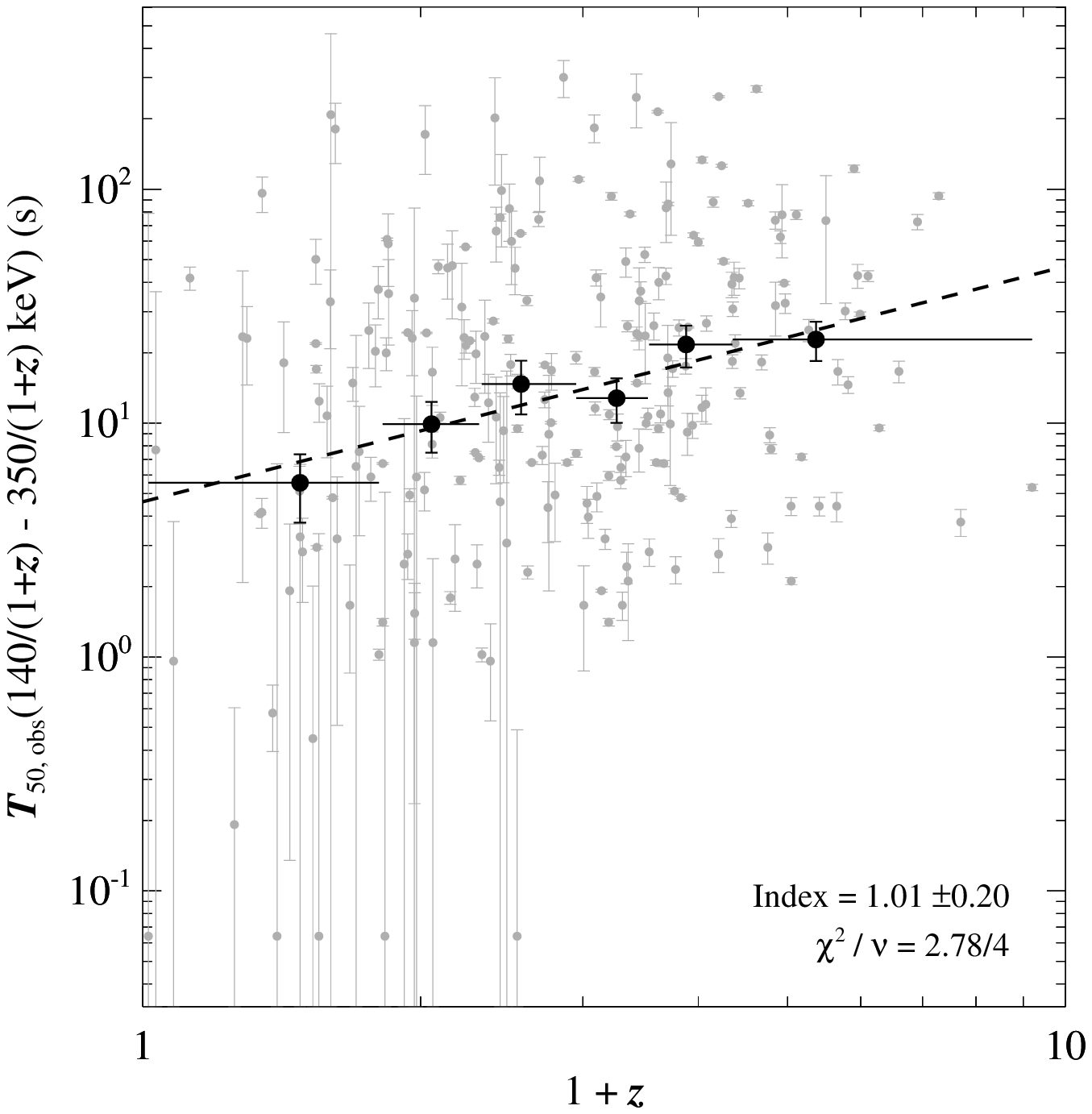}
    \\[-2ex]
    \includegraphics[width=7.5cm,height=7.2cm,angle=0,clip]{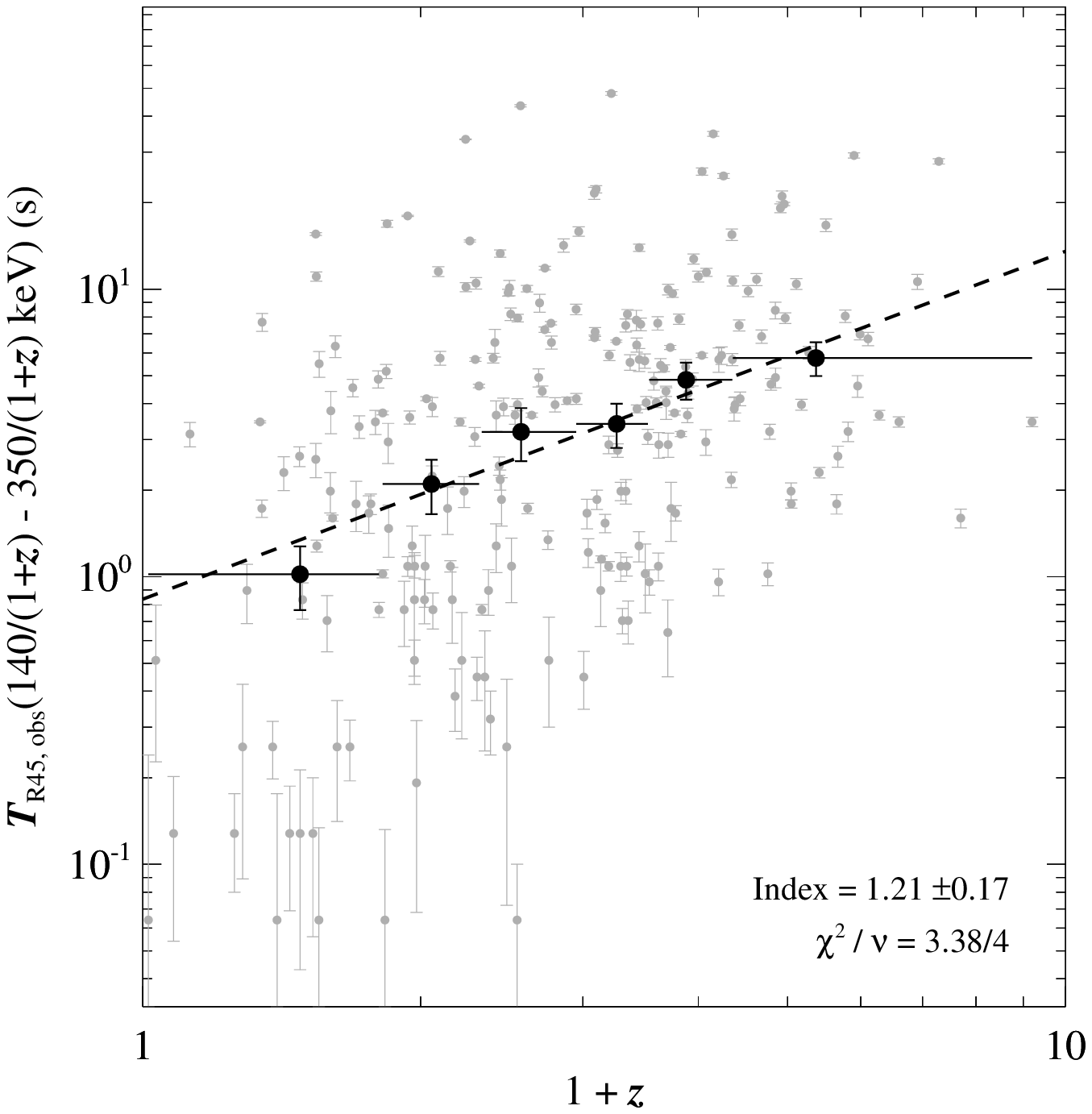}
    \\[-2ex]
    \caption{$T_{\rm 90,obs}$, $T_{\rm  50,obs}$ and $T_{\rm R45,obs}$
      obtained  in the $140/\left(1  + z  \right)$--$350/\left( 1  + z
      \right)$~keV  energy  range  for   the  232  GRBs  in  the  full
      \textit{Swift}/BAT  sample.   Grey   points  correspond  to  the
      individual   GRBs,  while   black  points   are   the  geometric
      average. In each panel the bins represented by the black average
      points contain the same GRBs.  The black dashed line corresponds
      to the  best fit power-law obtained when  modelling binned data.
      The $\chi^{2}$  fit statistic, degrees of  freedom and power-law
      index obtained are indicated in each panel.}
    \label{fig:full_bat_durs}
  \end{center}
\end{figure}

Having  determined  the full  and  bright  samples  to be  fitted,  we
considered  $T_{\rm 90,obs}$, $T_{\rm  50,obs}$ and  $T_{\rm R45,obs}$
for all  bursts. The full  distributions of all  three are shown  as a
function of redshift in Figure \ref{fig:full_bat_durs}, where the grey
points  are individual  GRBs and  the  black points  are average  bins
obtained using the geometric average of durations and redshifts within
each bin.\par

The three panels in  Figure \ref{fig:full_bat_durs} show that, for all
three duration  measures, the average duration tends  to increase with
redshift.  The final  average bin  of each  duration measure  does not
conform to this  trend, however. The fits shown in  each of the panels
are  also  detailed   in  Table  \ref{tab:ext_fit_details},  alongside
modelling of the bright subset for all three durations.\par

\begin{table*}
  \centering
  \caption{Details  of fits  to  geometric average  of  duration as  a
    function of  $\left( 1+z\right)$ for the extended  bright and full
    GRB  samples.   $N_{\rm  GRBs}$  is  the total  number  of  bursts
    contained in each fit, while  $\log_{10}N$ is the logarithm of the
    normalisation to each fitted power-law.}
  \label{tab:ext_fit_details}
  \begin{tabular}{cccccc}
    \hline
    \hline
    Sample & Duration & $N_{\rm GRBs}$ & $\log_{10}N$ & Index & $\chi^{2} / \nu$ \\
    \hline
    All & $T_{\rm 90,obs}$ & 232
    & 1.17 $\pm$0.13 & 0.83 $\pm$0.23 & 3.37/4 \\
    All & $T_{\rm 50,obs}$ & 232
    & 0.66 $\pm$0.11 & 1.01 $\pm$0.20 & 2.78/4 \\
    All & $T_{\rm RT45,obs}$ & 232
    & 0.08 $\pm$0.09 & 1.21 $\pm$0.17 & 3.38/4 \\
    \hline
    Bright & $T_{\rm 90,obs}$ & 89
    & 1.23 $\pm$0.17 & 0.64 $\pm$0.36 & 3.14/4 \\
    Bright & $T_{\rm 50,obs}$ & 89
    & 0.65 $\pm$0.14 & 0.77 $\pm$0.30 & 2.54/4 \\
    Bright & $T_{\rm RT45,obs}$ & 89
    & 0.32 $\pm$0.17 & 0.70 $\pm$0.37 & 5.83/4 \\
    \hline
  \end{tabular}
\end{table*}

For all three  durations, the behaviour of the  geometric average as a
function of redshift  can be fairly well described  by a power-law. In
all three cases,  however, the final average bin  is over-predicted by
the fitted power-law model.\par

Of    the    three    duration   measures,    $T_{\rm    50,obs}\left(
140/\left(1+z\right) -  350/\left(1+z\right)~{\rm keV}\right)$ has the lowest $\chi^{2}$ value, and  also  has  a value  most
consistent with that expected from time-dilation. Care must be taken,
though, as the $\chi^{2}$ fit  statistic indicates that the quality of
this fit is dominated by the statistical scatter of individual $T_{\rm
  50,obs}\left(   140/\left(1+z\right)   -   350/\left(1+z\right)~{\rm
  keV}\right)$ values within the bin.\par

As   shown   in   \citet{2013MNRAS.436.3640L},  at   high   redshifts,
difficulties arise in recovering periods of late-time pulse morphology
due to a  decreasing signal-to-noise ratio in the  pulse tail. $T_{\rm
  90}$ probes  the extended tail  of prompt emission more  deeply than
$T_{\rm 50}$,  and so  should be more  sensitive to these  effects. As
such,  if the  distribution  of  rest frame  GRB  durations is  indeed
constant,  then $T_{\rm  50,obs}$ might  be expected  to  most clearly
exhibit the  effects of cosmological  time dilation.  This is  what is
seen in  Figure \ref{fig:full_bat_durs}, with the  power-law fitted to
the geometric  average $T_{\rm 50,obs}$  as a function of  redshift is
closest to $\left( 1 + z \right)^{1}$.\par

Table  \ref{tab:ext_fit_details} also  indicates that  restricting the
sample to  only the brightest  bursts results in a  slightly shallower
power-law index.  With the exception of $T_{\rm R45,obs}$, this difference
between the  bright and full samples  for all three  durations is not,
however, greater than the error associated with the power-law index in
either fit.\par

We noted in Figure \ref{fig:full_bat_durs} that there are several GRBs
where  $\Delta  T_{\rm 90,obs}$  is  comparable  to,  or exceeds,  the
measured value  of $T_{\rm 90,obs}$.   We therefore re-fitted  both the
full and  bright updated \textit{Swift}/BAT sample  excluding all GRBs
where  $\Delta T_{\rm  90,obs} >  T_{\rm 90,obs}$.   This  reduced the
sample sizes to 220 and  84, respectively.  When filtering the data in
this way, the GRBs that were removed tended to be in the low redshift,
low    duration    regions   of    our    parameter   space.     Table
\ref{tab:weed_fit_details}  details  the   fitted  power-laws  to  the
geometric mean average bins for  each duration after the data had been
filtered.\par

\begin{table*}
  \centering
  \caption{Details  of fits  to  geometric average  of  duration as  a
    function of  $\left( 1+z\right)$ for the  filtered extended bright
    and  full GRB  samples.  $N_{\rm  GRBs}$  is the  total number  of
    bursts considered in each fit, while $\log_{10}N$ is the logarithm
    of the normalisation to each fitted power-law.}
  \label{tab:weed_fit_details}
  \begin{tabular}{cccccc}
    \hline
    \hline
    Sample & Duration & $N_{\rm GRBs}$ & $\log_{10}N$ & Index & $\chi^{2} / \nu$ \\
    \hline
    All & $T_{\rm 90,obs}$ & 220
    & 1.39 $\pm$0.12 & 0.51 $\pm$0.23 & 4.16/4 \\
    All & $T_{\rm 50,obs}$ & 220
    & 0.87 $\pm$0.08 & 0.69 $\pm$0.16 & 2.04/4 \\
    All & $T_{\rm RT45,obs}$ & 220
    & 0.08 $\pm$0.09 & 1.01 $\pm$0.18 & 4.26/4 \\
    \hline
    Bright & $T_{\rm 90,obs}$ & 84
    & 1.37 $\pm$0.22 & 0.40 $\pm$0.49 & 6.20/4 \\
    Bright & $T_{\rm 50,obs}$ & 84
    & 0.81 $\pm$0.21 & 0.47 $\pm$0.48 & 7.70/4 \\
    Bright & $T_{\rm RT45,obs}$ & 84
    & 0.42 $\pm$0.17 & 0.53 $\pm$0.37 & 7.88/4 \\
    \hline
  \end{tabular}
\end{table*}

In  all instances,  as shown  in Tables  \ref{tab:ext_fit_details} and
\ref{tab:weed_fit_details},   the  power-law   index   decreases  when
removing  those bursts  with large  relative uncertainties  in $T_{\rm
  90,obs}$.   Comparing  these  new  values with  that  expected  from
cosmological time dilation, we find that only $T_{\rm R45,obs}$ now is
consistent with this hypothesis. However, it is important to note that
when   sampling   only   the   brightest   bursts,   as   defined   by
\citet{2013ApJ...778L..11Z}, the correlation of $T_{\rm R45,obs}$ with
redshift is shallower and is more poorly described by a power-law.\par

Removing bursts with greater  relative uncertainty in $T_{\rm 90,obs}$
demonstrates that  care must  be taken when  defining the  sample over
which any of  these putative correlations are to  be measured. In this
instance, the  correlations favour fitted values  more consistent with
cosmological  time dilation  when  also considering  GRBs with  poorly
defined values of duration.\par

Modelling  the evolution  of redshift  in the  manner  described above
allows us to compare the geometric average value of a duration measure
as a  function of redshift,  however, it does not  provide information
regarding  the  shape  of  the distribution.   We  therefore  compared
subsets of the duration distributions, defined by redshift.\par

We  calculated the  rest  frame values  of  $T_{\rm 90,obs}$,  $T_{\rm
  50,obs}$  and  $T_{\rm R45,obs}$  (where  e.g.   $T_{\rm 90,rest}  =
T_{\rm  90,obs}/ \left(  1 +  z \right)$),  which correspond  to being
measured in the  140--350~keV range of the rest frame  of each GRB. We
then isolated four subsets within the full extended sample: those with
a redshift above the median  redshift of the sample, $z_{\rm median} =
1.95$;  those with $z  < z_{\rm  median}$; GRBs  with redshift  in the
upper  quartile of  the sample  distribution and  finally GRBs  with a
redshift in the lower quartile of the sample distribution.\par

\begin{figure*}
  \begin{center}
    \includegraphics[width=7.5cm,height=7.2cm,clip,angle=0]{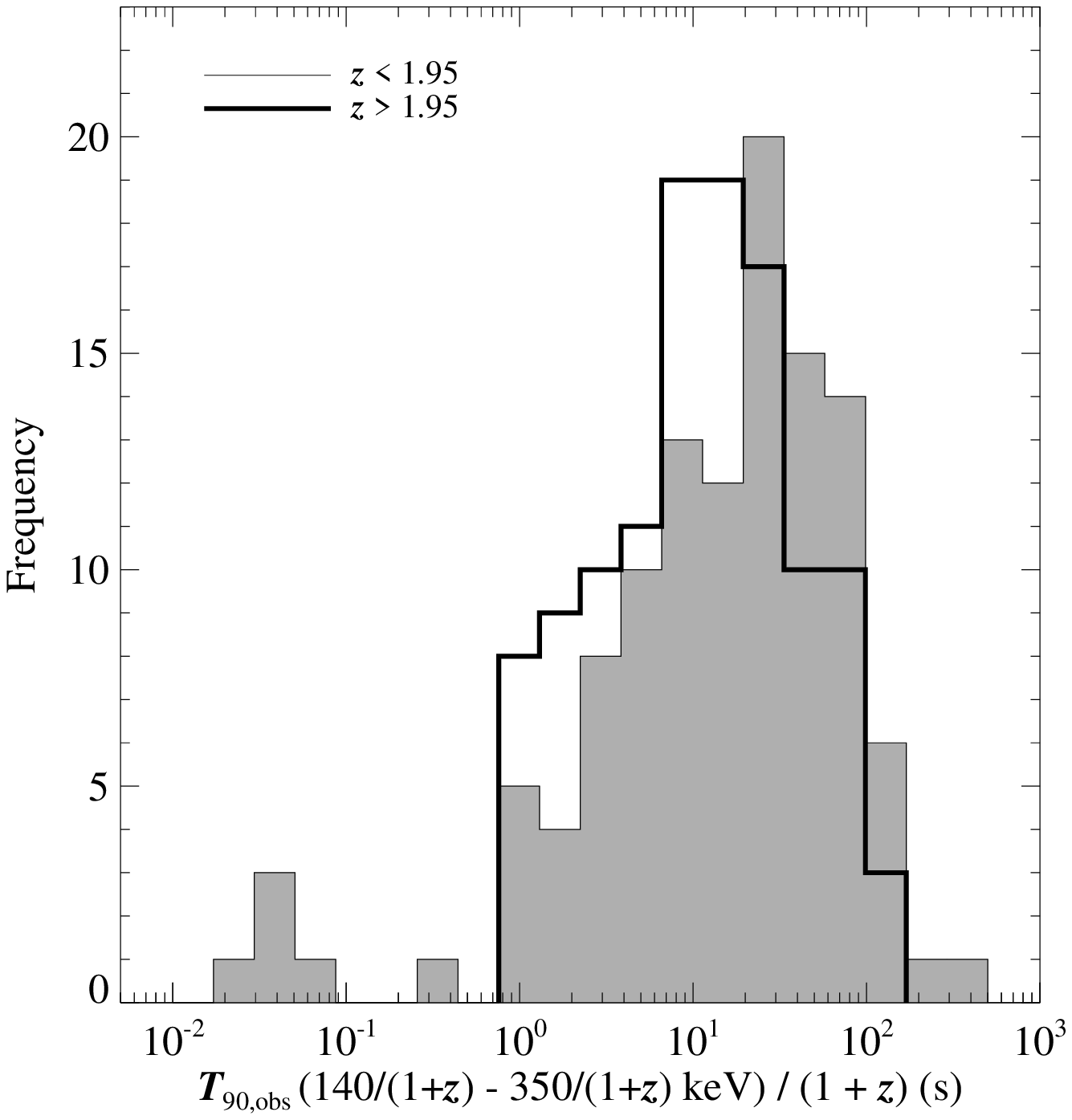}
    \quad
    \includegraphics[width=7.5cm,height=7.2cm,clip,angle=0]{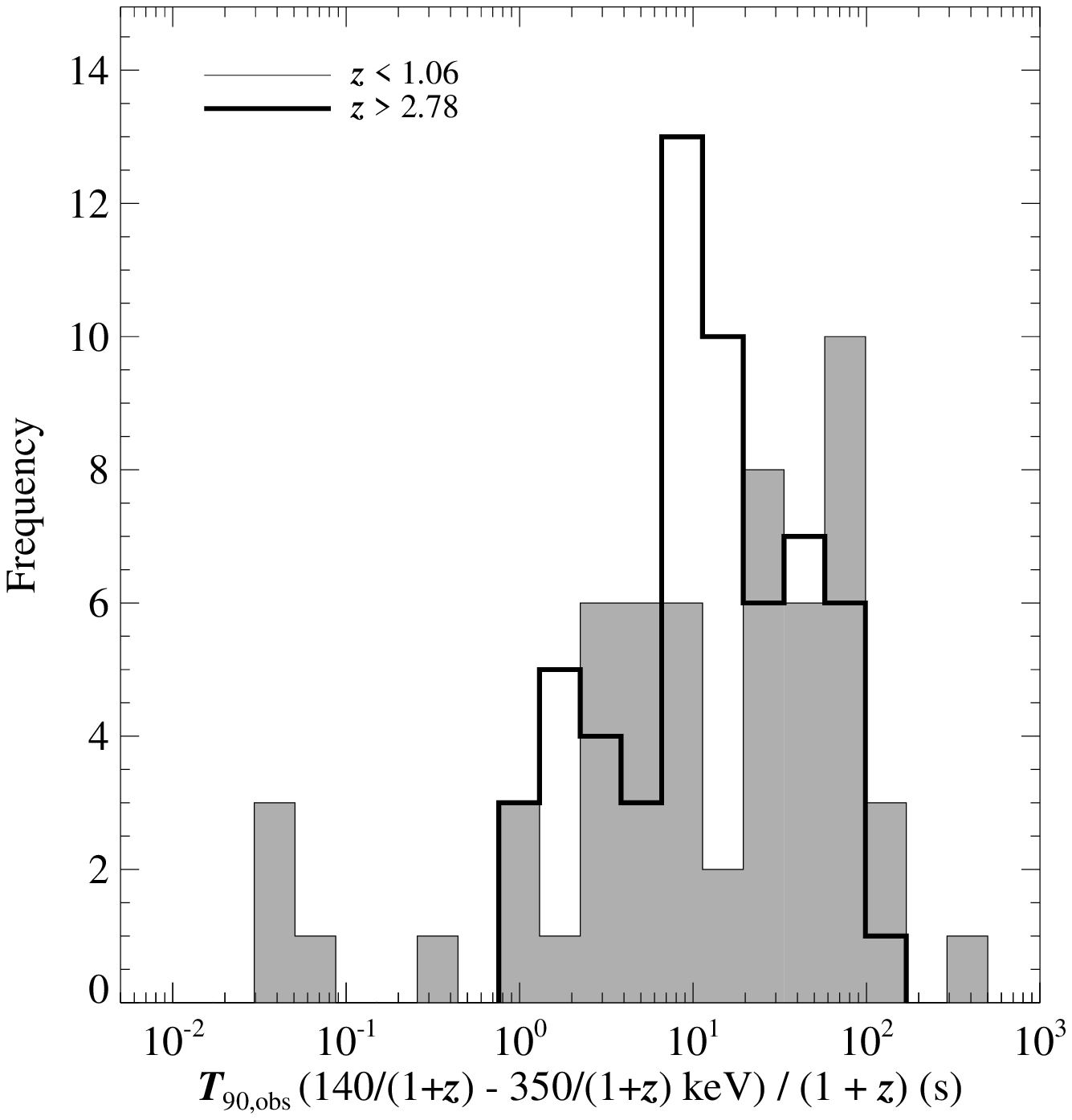}
    \\[-2ex]
    \includegraphics[width=7.5cm,height=7.2cm,clip,angle=0]{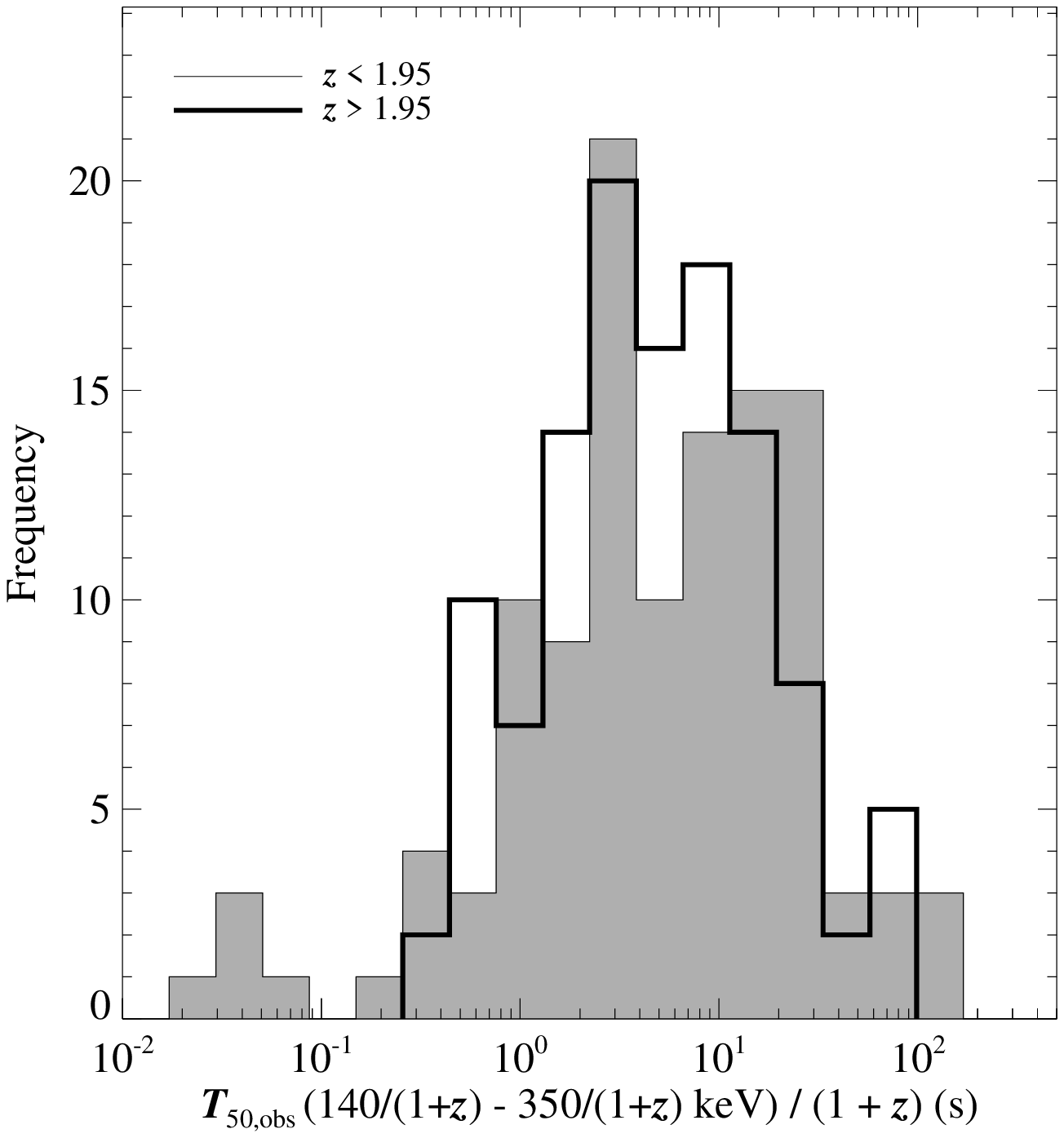}
    \quad
    \includegraphics[width=7.5cm,height=7.2cm,clip,angle=0]{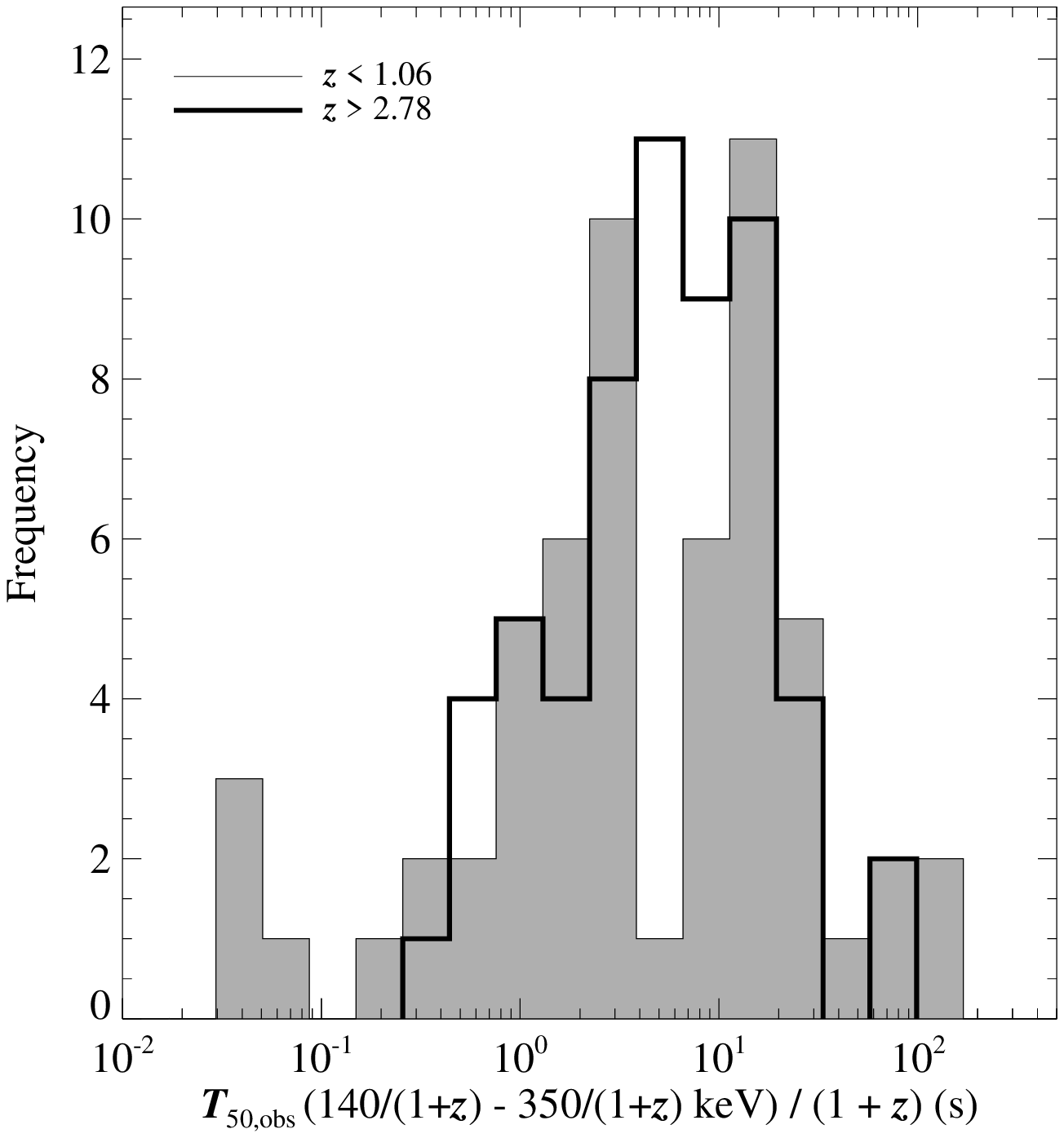}
    \\[-2ex]
    \includegraphics[width=7.5cm,height=7.2cm,clip,angle=0]{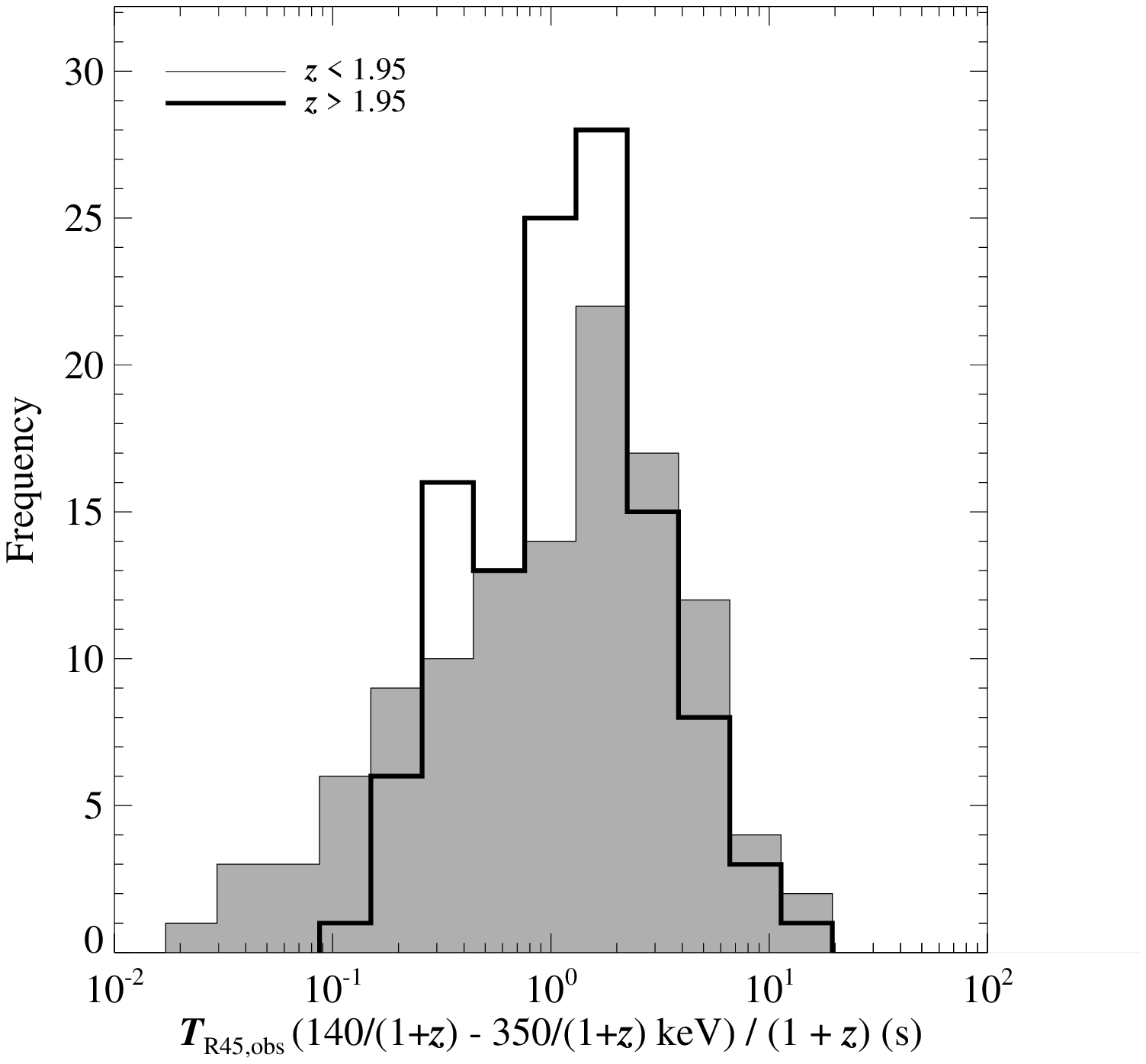}
    \quad
    \includegraphics[width=7.5cm,height=7.2cm,clip,angle=0]{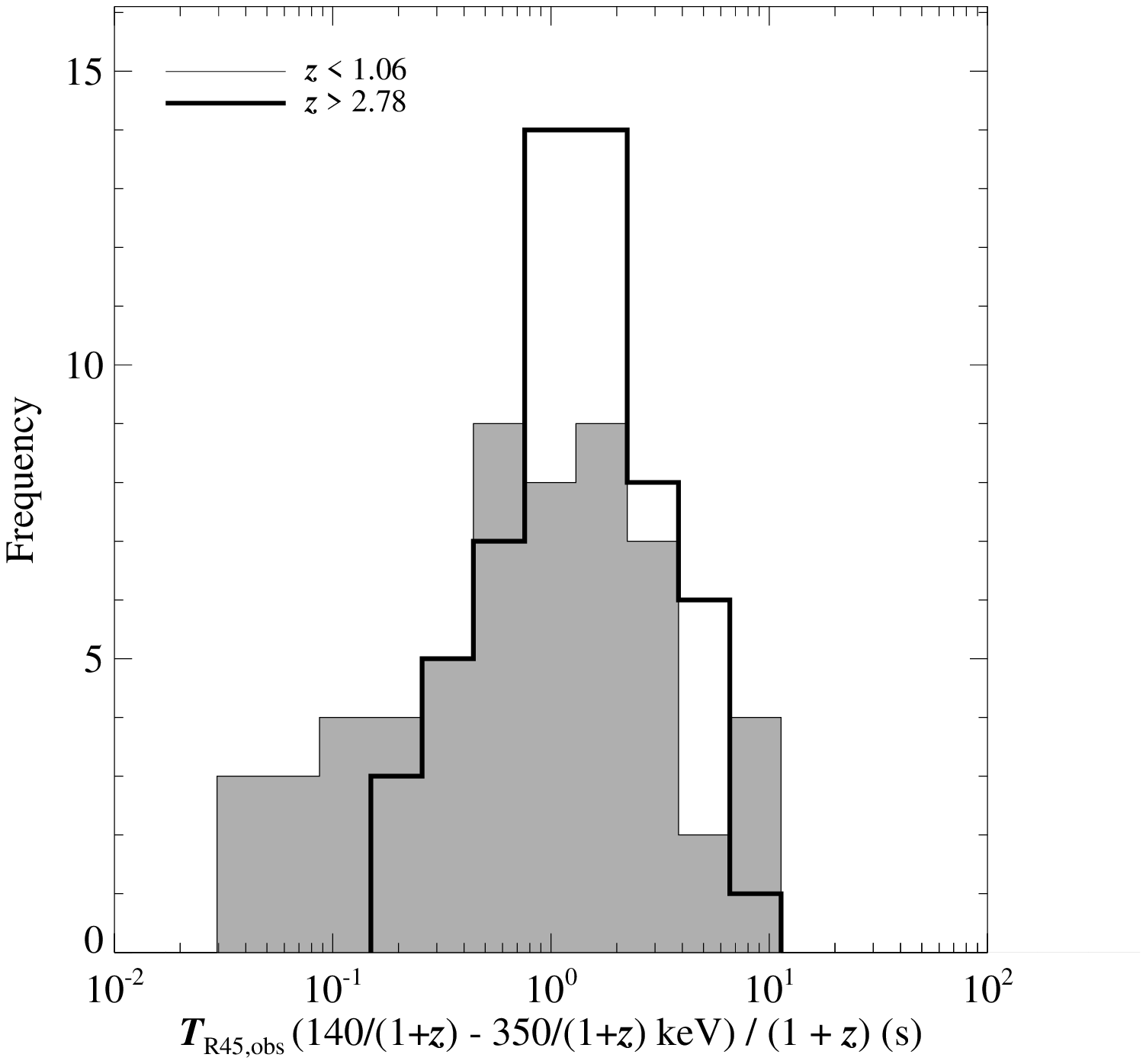}
    \\[-2ex]
  \end{center}
  \caption{Histograms comparing rest frame $\left( 140~-~350~{\rm keV}
    \right)$ durations for sub-samples of the GRB population.  The left
    panels compare  the 116 GRBs  with $z <  z_{\rm median}$ to  the 116
    with $z  > z_{\rm median}$. The  right hand panel  compares the 58
    bursts  in the  lowest redshift  quartile to  the 58  GRBs  in the
    highest  redshift quartile.  The  top panels  are comparisons  of
    $T_{\rm 90,rest}$, the middle two panels are $T_{\rm 50,rest}$ and
    the bottom two are $T_{\rm R45,rest}$.}
  \label{fig:rest_frame_dists}
\end{figure*}

We then performed a Student's $t$-test comparing GRBs with $z > z_{\rm
  median}$  to those  with  $z  < z_{\rm  median}$  for each  duration
measure. As the duration  measures are best represented in logarithmic
space,    these     Student's    $t$-tests    were     performed    using
$\log_{10}\left(T_{\rm dur,rest}\right)$, where $T_{\rm dur,rest}$ was
$T_{\rm 90,rest}$,  $T_{\rm 50,rest}$ or $T_{\rm  R45,rest}$.  We also
compared the duration of bursts with redshift in the upper quartile to
the durations  of bursts  with redshifts in  the lower  quartile.  The
distributions   of  rest   frame   durations  are   shown  in   Figure
\ref{fig:rest_frame_dists}.   The   results  for  each   $t$-test  are
detailed in Table \ref{tab:t_tests}.\par

\begin{table}
  \centering
  \caption{Results  from  Student's  $t$-tests  comparing  rest  frame
    duration measures. $N_{1}$ and $N_{2}$  give the number of GRBs in
    each subset,  $z_{1}$ and $z_{2}$  detail the redshift  limits for
    each  distribution. $t$  is the  Student's $t$-test  statistic and
    $p\left(H_{0}\right)$   is   the    probability   that   the   two
    distributions have the same mean.}
  \label{tab:t_tests}
  \begin{tabular}{ccccccc}
    \hline
    \hline
    Duration & $N_{1}$ & $N_{2}$ & $z_{1}$ & $z_{2}$ & $t$ & $p\left(H_{0}\right)$ \\
    \hline
    $T_{\rm 90,rest}$ & 116 & 116 & $<$1.95 & $>$1.95 & 0.81 & 0.42 \\
    $T_{\rm 50,rest}$ & 116 & 116 & $<$1.95 & $>$1.95 & 0.15 & 0.88 \\
    $T_{\rm R45,rest}$ & 116 & 116 & $<$1.95 & $>$1.95 & -1.11 & 0.27 \\
    \hline
    $T_{\rm 90,rest}$ & 58 & 58 & $<$1.06 & $>$2.78 & -0.22 & 0.82 \\
    $T_{\rm 50,rest}$ & 58 & 58 & $<$1.06 & $>$2.78 & -0.70 & 0.47 \\
    $T_{\rm R45,rest}$ & 58 & 58 & $<$1.06 & $>$2.78 & -2.47 & 0.01 \\
    \hline
  \end{tabular}
\end{table}

From Figure \ref{fig:rest_frame_dists}  and Table \ref{tab:t_tests} it
can be  seen that for five of  the six statistical tests,  there is no
significant  difference between the  populations being  compared. This
supports   the   claims   of  \citet{2013ApJ...778L..11Z}   that   the
distributions of prompt emission rest  frame durations in a rest frame
defined energy band are constant.\par

The results of  the test comparing those values  of $T_{\rm R45,rest}$
in  the  lower  redshift  quartile  to those  in  the  upper  redshift
quartile,  however, indicate  a difference  in the  distribution  at a
significance of  approximately 3$\sigma$.   The sixth panel  of Figure
\ref{fig:rest_frame_dists} shows  this is because  the distribution at
higher redshifts appears to be  narrower and, in particular, lacks low
values of $T_{\rm  R45,rest}$. It is worth noting,  as shown in Figure
\ref{fig:full_bat_durs}  that  these low  duration  values of  $T_{\rm
  R45,obs}$  have   higher  measured  errors,   indicating  a  greater
uncertainty in these values.\par

$T_{\rm  R45,obs}$  fundamentally differs  from  $T_{\rm 90,obs}$  and
$T_{\rm 50,obs}$,  as the former  probes only the brightest  region of
prompt emission. As such, $T_{\rm R45,obs}$ is more insensitive to the
presence  quiescent periods of  a light  curve.  Conversely,  should a
quiescent  period occur  within the  central 50\%  of  GRB high-energy
fluence, both $T_{\rm 90,obs}$ and $T_{\rm 50,obs}$ would include that
period. It  is perhaps expected, therefore, that  $T_{\rm R45,obs}$ is
less likely to exhibit evidence of cosmological time dilation.\par

While  most pronounced  in  the $T_{\rm  R45}$ distributions,  $T_{\rm
  90,rest}$ and $T_{\rm 50,rest}$ also  show a small population of low
duration, low redshift  bursts. It is possible that  these are fainter
GRBs,  and as  such cannot  be detected  by the  \textit{Swift}/BAT at
higher  redshifts.  This effect  is  investigated  in  more detail  in
\S~\ref{sec:hi_z_short_dur}.\par

\subsection{Durations measured by \textit{Fermi}/GBM}
\label{sec:t90_gbm}

By  extracting durations  in a  rest frame  defined energy  range, the
observer energy  range is  a function of  redshift. As  such, detector
effects  may  inhomogeneously alter  the  recovered  durations of  the
sample.   To  test  whether   such  effects   are  present,   we  also
cross-referenced  the  232 GRBs  in  our  extended  sample with  those
observed  by \textit{Fermi}/GBM.   This yielded  57 GRBs  for  which a
\textit{Swift}/BAT  light curve,  \textit{Fermi}/GBM light  curves and
redshift measurement were available.\par

The \textit{Fermi}/GBM  $140/\left( 1 + z \right)$--$350/\left(  1 + z
\right)$~keV light curves were extracted  for each GRB. We then binned
the    light    curves   using    the    methodology   discussed    in
\S~\ref{sec:zhang_chk}.   We also  took  the \textit{Swift}/BAT  light
curves for only this subset of 57 GRBs to ensure these burst durations
were representative of the larger \textit{Swift}/BAT sample.\par

\begin{figure*}
  \begin{center}
    \includegraphics[width=7.5cm,height=7.2cm,angle=0,clip]{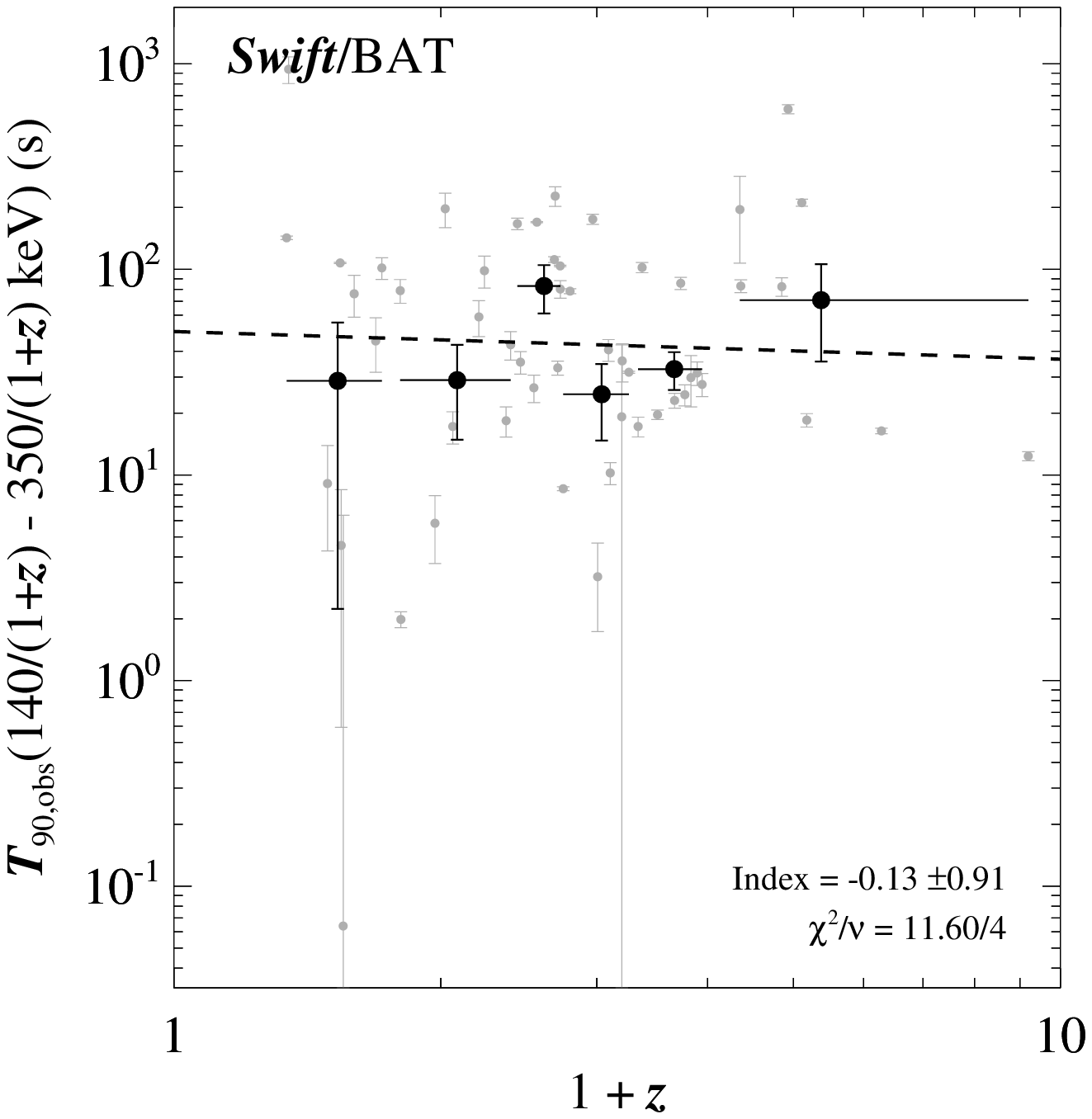}
    \quad
    \includegraphics[width=7.5cm,height=7.2cm,angle=0,clip]{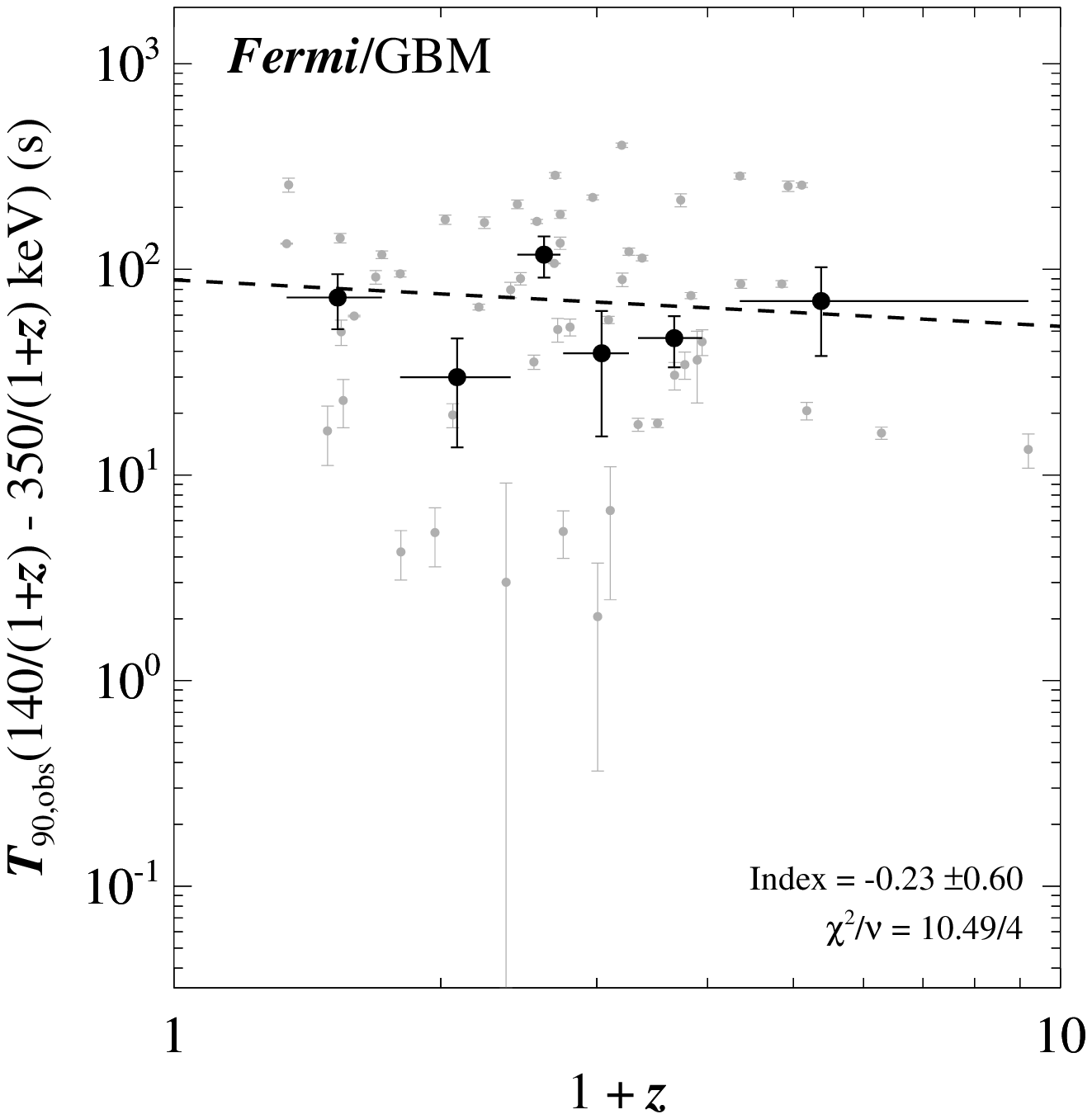}
    \\[-2ex]
    \includegraphics[width=7.5cm,height=7.2cm,angle=0,clip]{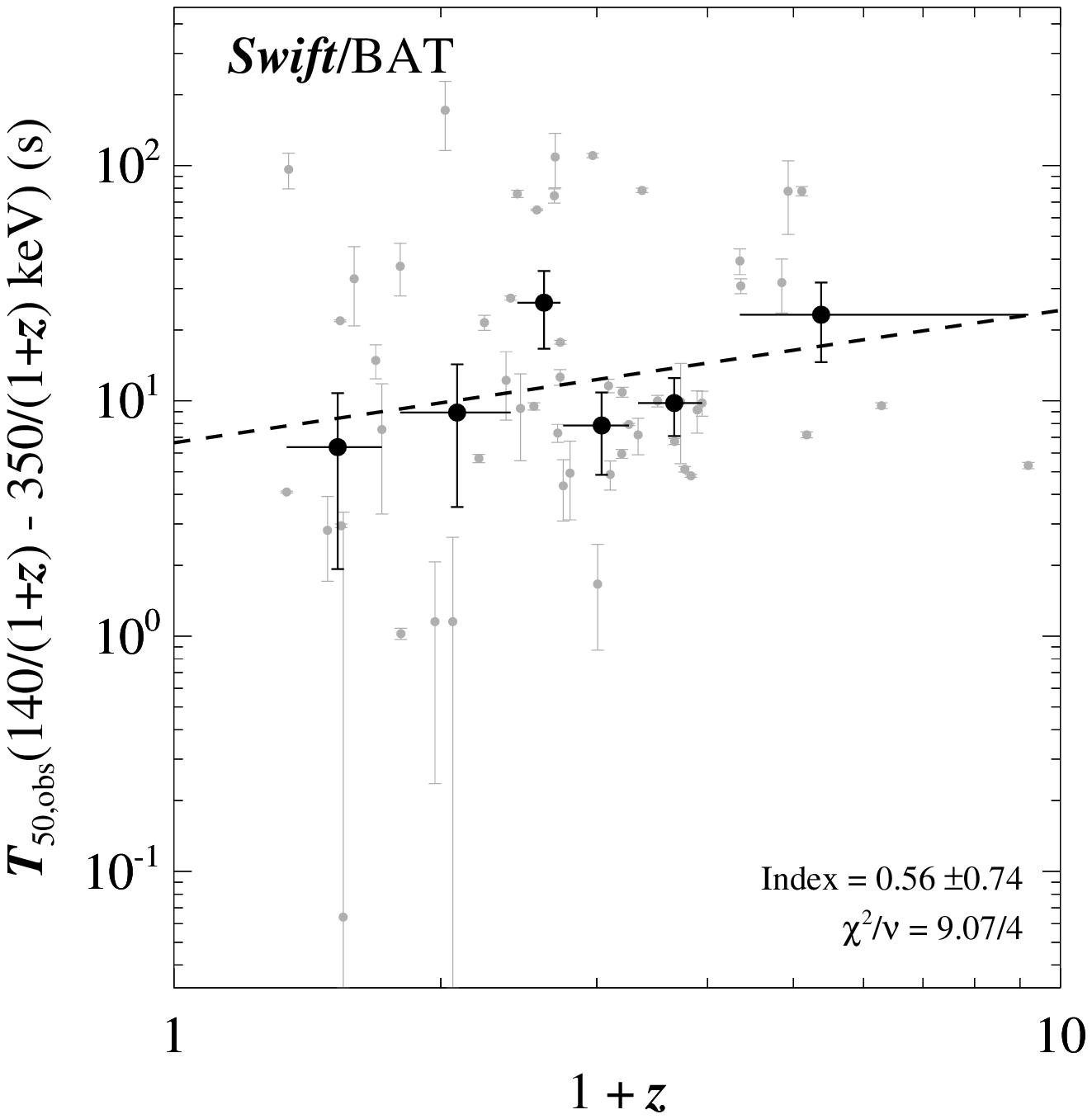}
    \quad
    \includegraphics[width=7.5cm,height=7.2cm,angle=0,clip]{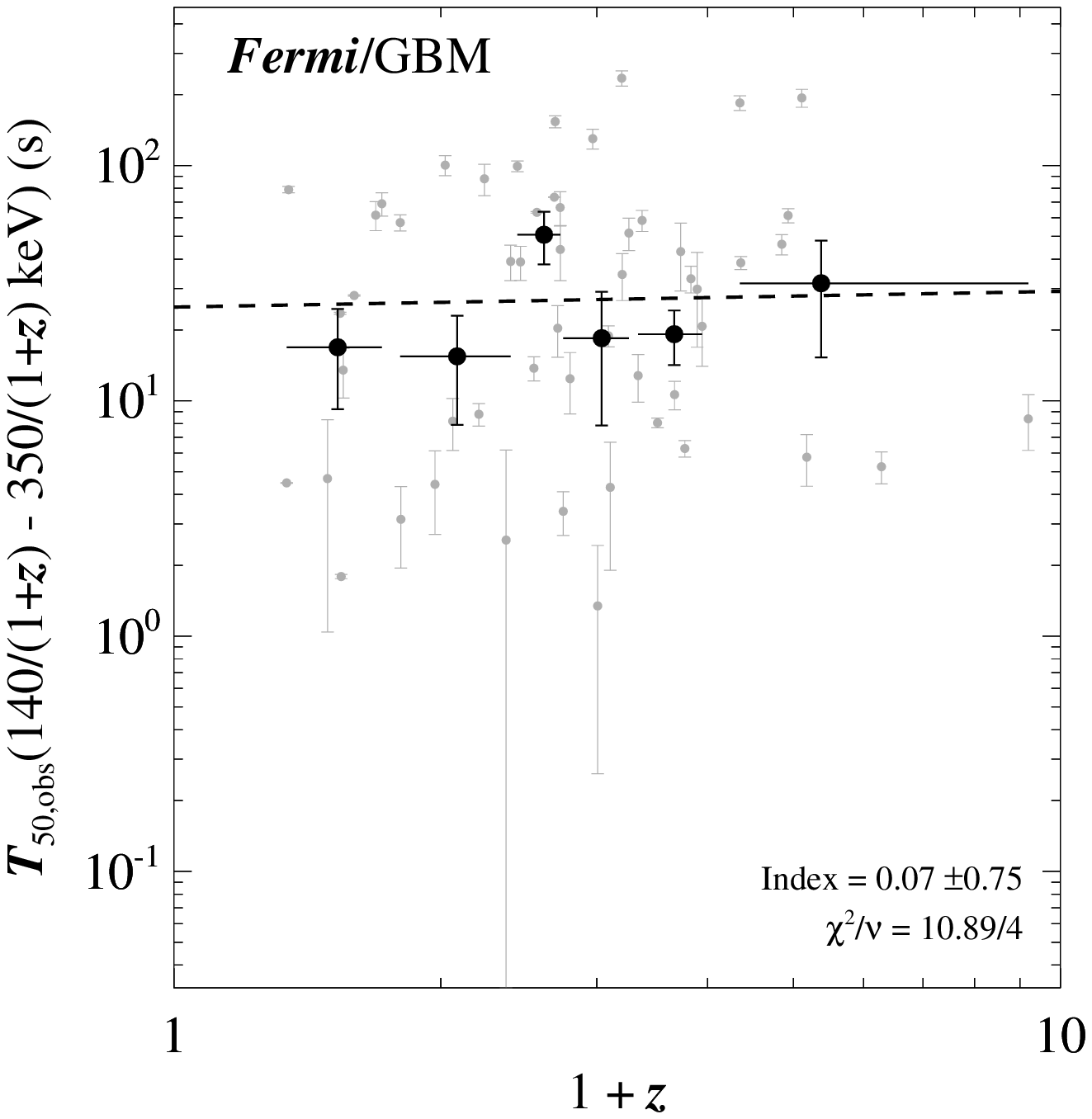}
    \\[-2ex]
    \includegraphics[width=7.5cm,height=7.2cm,angle=0,clip]{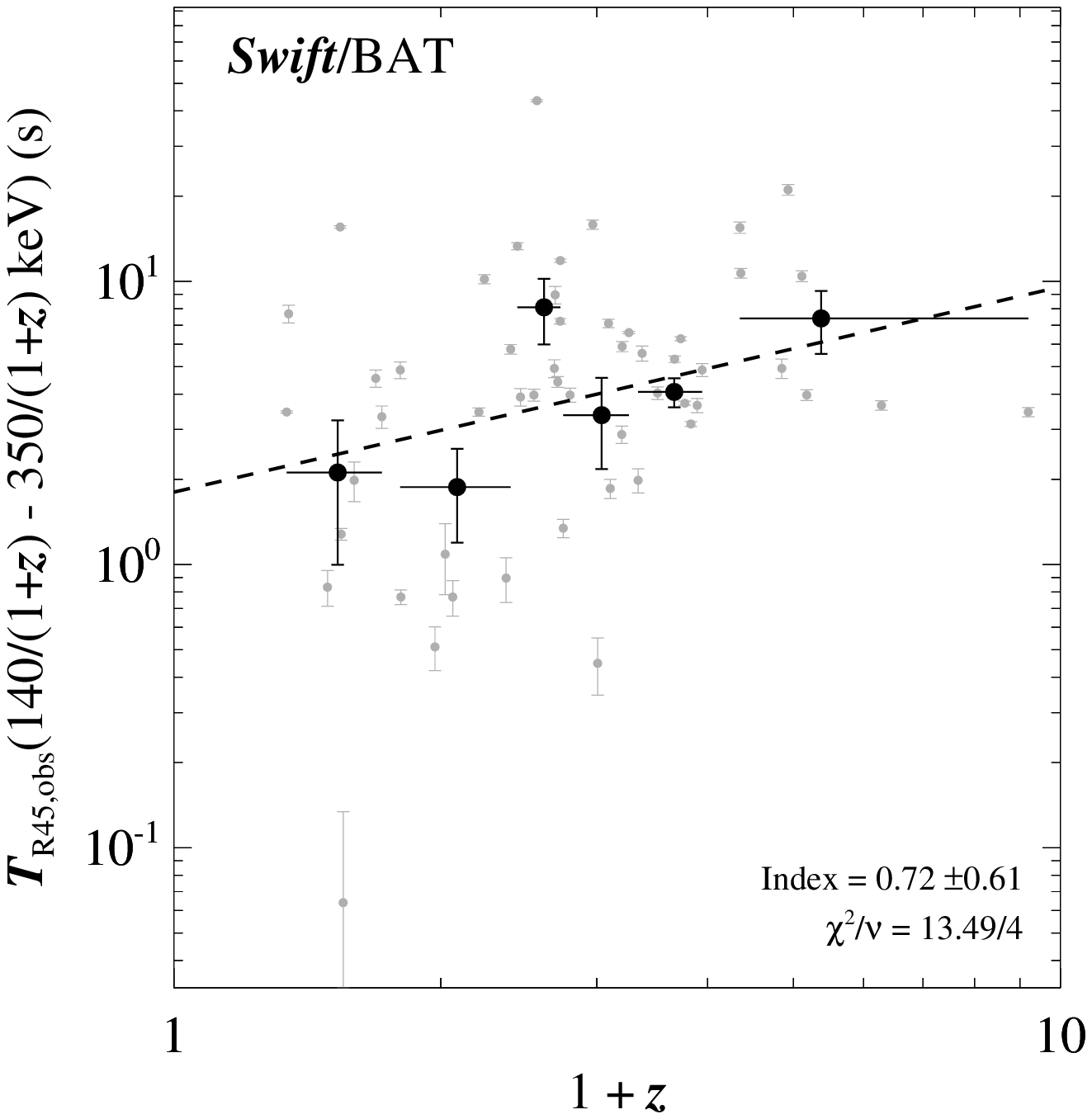}
    \quad
    \includegraphics[width=7.5cm,height=7.2cm,angle=0,clip]{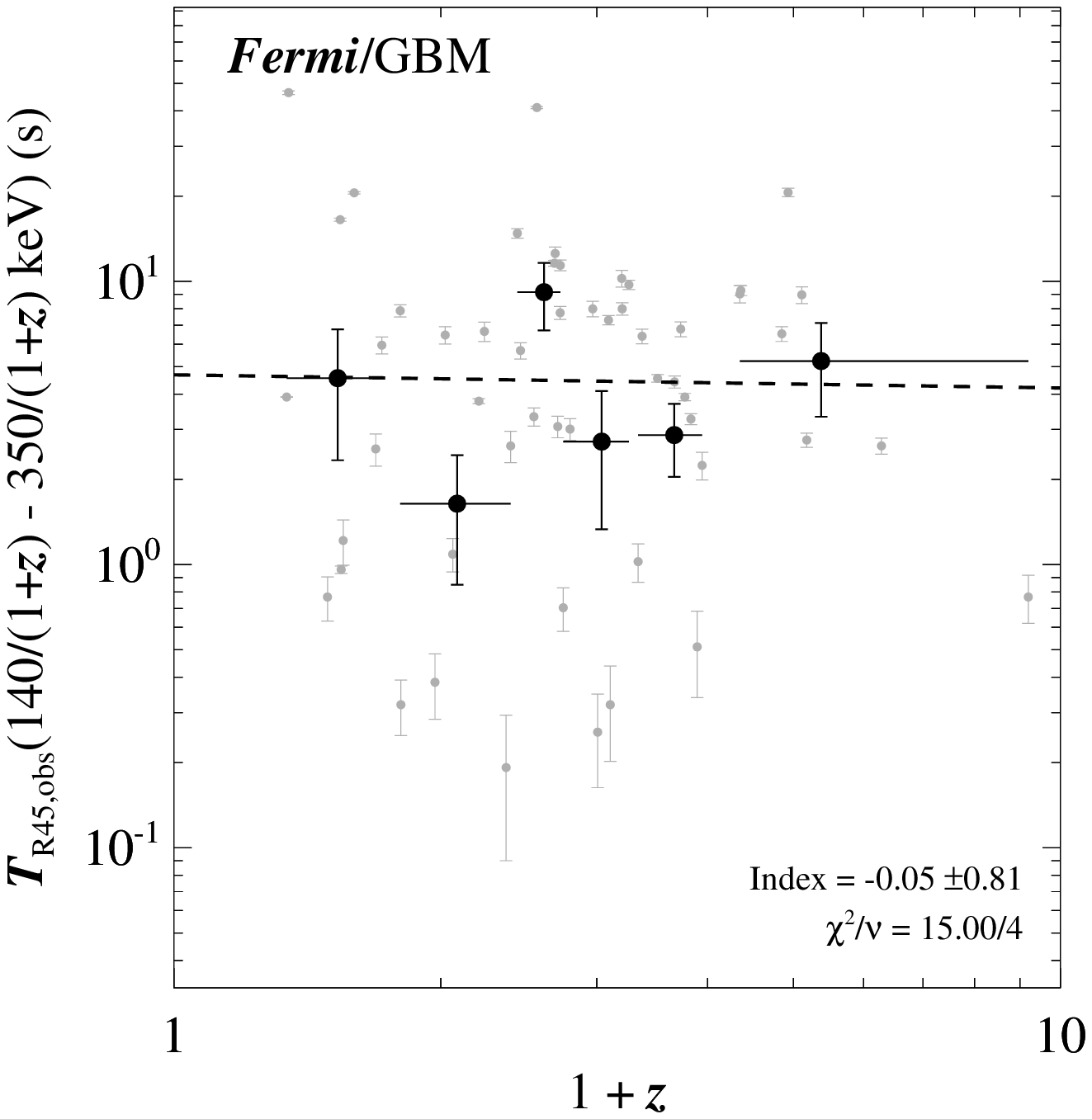}
    \\[-2ex]
  \end{center}
  \caption{$T_{\rm  90,obs}$, $T_{\rm  50,obs}$ and  $T_{\rm R45,obs}$
    obtained  in the  $140/\left(1  + z  \right)$--$350/\left(  1 +  z
    \right)$~keV  energy   range  for  the   57  GRBs  in   the  joint
    \textit{Swift}/BAT-\textit{Fermi}/GBM    sample.     Grey   points
    correspond  to the  individual GRBs,  while black  points  are the
    geometric  average.  \textit{Swift}/BAT  data  are  shown  in  the
    left-hand column, while  the corresponding \textit{Fermi}/GBM data
    are shown in the right column. Binning of GRBs is identical in all
    six panels. The  black dashed line in each  panel corresponds to a
    power-law  fitted  to  the   average  bins.   The  $\chi^{2}$  fit
    statistics, degrees  of freedom and  power-law indices of  all six
    models are indicated in each panel.}
  \label{fig:all_durs}
\end{figure*}

Figure  \ref{fig:all_durs}  shows  the  individual  duration  measures
obtained from both the \textit{Swift}/BAT and \textit{Fermi}/GBM light
curves in  grey. Also shown in  black are the geometric  averages as a
function  of redshift. The  binning of  GRBs is  identical in  all six
panels.  Details   of  the  fitted  power-laws  are   given  in  Table
\ref{tab:gbm_fits}.\par

\begin{table}
  \centering
  \caption{Fitted power-law  parameters and fit statistics  for the 57
    GRB  joint \textit{Swift}/BAT-\textit{Fermi}/GBM subset.  For each
    fit,  the  duration considered  and  instrument  from which  light
    curves were taken are shown.  $\log_{10}N$ is the logarithm of the
    normalisation to each fitted power-law.}
  \label{tab:gbm_fits}
  \begin{tabular}{ccccc}
    \hline
    \hline
    Instrument & Duration & $\log_{10}N$ & Index & $\chi^{2} / \nu$ \\
    \hline
    \textit{Swift}/BAT & $T_{\rm 90,obs}$ & 1.70 $\pm$0.47 & -0.13 $\pm$0.91 
    & 11.60/4 \\
    \textit{Swift}/BAT & $T_{\rm 50,obs}$ & 0.82 $\pm$0.40 & 0.56 $\pm$0.74 
    & 9.07/4 \\
    \textit{Swift}/BAT & $T_{\rm R45,obs}$ & 0.26 $\pm$0.34 & 0.72 $\pm$0.61 
    & 13.49/4 \\
    \hline
    \textit{Fermi}/GBM & $T_{\rm 90,obs}$ & 1.95 $\pm$0.27 & -0.23 $\pm$0.60 
    & 10.49/4 \\
    \textit{Fermi}/GBM & $T_{\rm 50,obs}$ & 1.40 $\pm$0.36 & 0.07 $\pm$0.75 
    & 10.89/4 \\
    \textit{Fermi}/GBM & $T_{\rm R45,obs}$ & 0.67 $\pm$0.41 & -0.05 $\pm$0.81 
    & 15.00/4 \\
    \hline
  \end{tabular}
\end{table}

Figure \ref{fig:all_durs} appears to  show that the geometric averages
of  all three duration  measures are  less positively  correlated with
redshift when calculated for  the same sample using \textit{Fermi}/GBM
light  curves.  Inspection of  Figure  \ref{fig:all_durs} reveals  the
duration distributions for  both instruments differ most significantly
at low redshifts.\par

However, with a large intrinsic scatter in the duration distributions,
the geometric  average bins are  less well represented by  a power-law
model fit.   Due to the large  scatter of duration  values within each
bin the statistical  errors of these bins, and  subsequently the model
fit parameters,  are large. The differences in  power-law index fitted
to all three duration distributions  is approximately the same size as
the error in  the fitted parameter. This indicates  a larger sample of
\textit{Fermi}/GBM GRBs  with measured redshift is  required to assess
the  significance  of  this  difference.  Comparisons  between  Tables
\ref{tab:weed_fit_details}  and \ref{tab:gbm_fits}  also  show that  a
larger  sample is  required  to  be more  representative  of the  full
\textit{Swift}/BAT  sample  and remove  the  effects  of small  number
statistics.\par

\section{Discussion}
\label{sec:disc}

\subsection{Low-redshift long-duration GRBs}
\label{sec:lo_z_long_dur}

There are  two regions of the redshift-duration  parameter space that,
if  artificially underpopulated, would  enhance a  correlation between
the  two  quantities.   The  first  is  at  low  redshifts,  but  long
durations.  At low-redshifts, the  sample of  observed GRBs  is volume
limited.  The observed GRB  sample should  therefore reflect  the most
common type of burst within the total distribution.\par

A volume limited  sample of GRBs should also contain  GRBs of the most
common intrinsic  luminosities.  Previous studies  have suggested that
the luminosity function of GRBs can be described by either a single of
broken                                                        power-law
\citep{2012ApJ...749...68S,2011MNRAS.416.2174C,2010ApJ...711..495B},
such  that  more bursts  are  expected from  the  fainter  end of  the
luminosity distribution.   For faint  GRBs any late-time  prompt light
curve morphology  would remain undetected due  to poor signal-to-noise
ratio \citep{2013MNRAS.436.3640L}.\par

At the  lowest redshifts  ($z \leqslant 0.5$),  it must also  be noted
that the $140/\left(1 + z \right)-350/\left(1+ z \right)$ energy range
samples a region of the \textit{Swift}/BAT response with significantly
lower  effective area.  This  will further  reduce the  total observed
burst   sample,  but  in   a  uniform   manner  across   the  duration
distribution.\par

Considering the $T_{\rm 90,rest}$ distribution of all 232 long GRBs we
find  41 bursts  with a  rest frame  duration $T_{\rm  90,rest}  > 50$
seconds.   This   corresponds  to  a   probability  of  $p\left(T_{\rm
  90,rest}> 50{\rm ~s}\right) =  0.18$. Considering the GRBs in
the lowest quartile of redshifts  ($z < 1.06$), we find $p\left(T_{\rm
  90,rest}>50{\rm ~s},~z  < 1.06\right)  = 0.25$  (15/59). As
such,  there is no  evidence that  this region  of parameter  space is
under-sampled.\par

\subsection{The lack of high-redshift shorter-duration GRBs}
\label{sec:hi_z_short_dur}

One region of  parameter space that is clearly  under-populated in all
panels  of   Figure  \ref{fig:full_bat_durs}  is   the  high-redshift,
short-duration  quadrant.   It is  important  to  note  that, in  this
instance, the  term short is relative  to the rest  of the population,
and  as  such all  bursts  considered  still  ascribe to  the  classic
definition of long GRBs \citep{1993ApJ...413L.101K}.\par

Referring     back    to     Figures     \ref{fig:full_bat_durs}    and
\ref{fig:rest_frame_dists},  we   postulated  that  bursts   with  low
durations  at low redshifts  are fainter,  and therefore  difficult to
detect. A  consequence of this would be  that \textit{Swift}/BAT would
become unable to detect the same population of GRBs if present at high
redshifts, due to the reduced signal-to-noise ratio.\par

To investigate whether this lack of high-redshift, short-duration GRBs
is a consequence of an  inability of \textit{Swift}/BAT to detect such
bursts, we  considered the total  signal-to-noise ratio of  the prompt
light  curves.  To  do this,  we de-noised  the  background subtracted
$15-350$~keV      light      curves      using      Haar      wavelets
\citep{2002A&A...385..377Q,1997ApJ...483..340K}. We then integrated to
total  smoothed fluence per  detector (in  counts) within  the $T_{\rm
  90,obs}$ duration for each light curve.  The full \textit{Swift}/BAT
energy range  was selected as this  is the most  indicative of whether
\textit{Swift}/BAT would trigger on a given GRB.  \par

\begin{figure}
  \begin{center}
    \includegraphics[width=7.5cm,height=7.2cm,angle=0,clip]{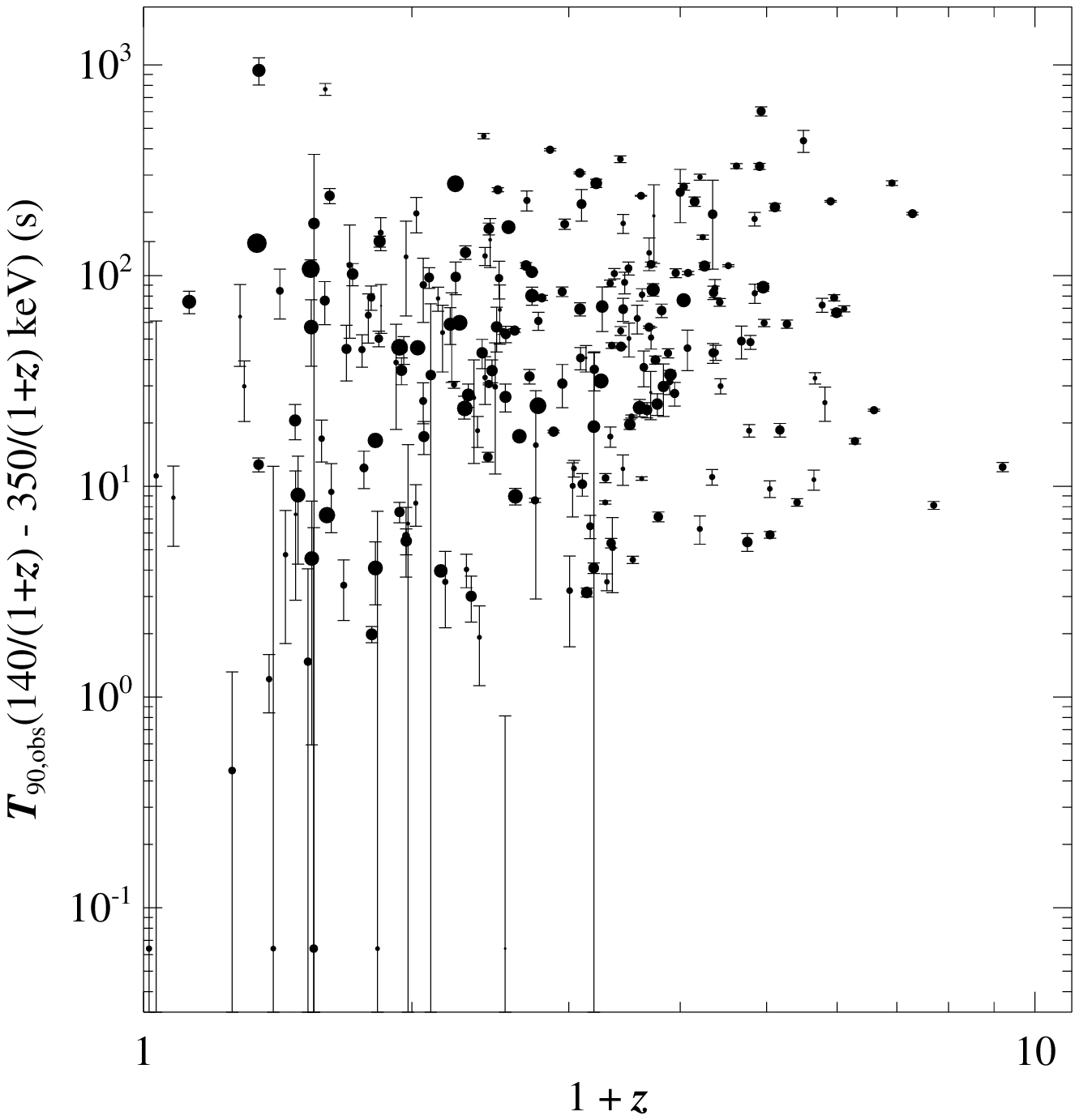}
    \\[-2ex]
    \includegraphics[width=7.5cm,height=7.2cm,angle=0,clip]{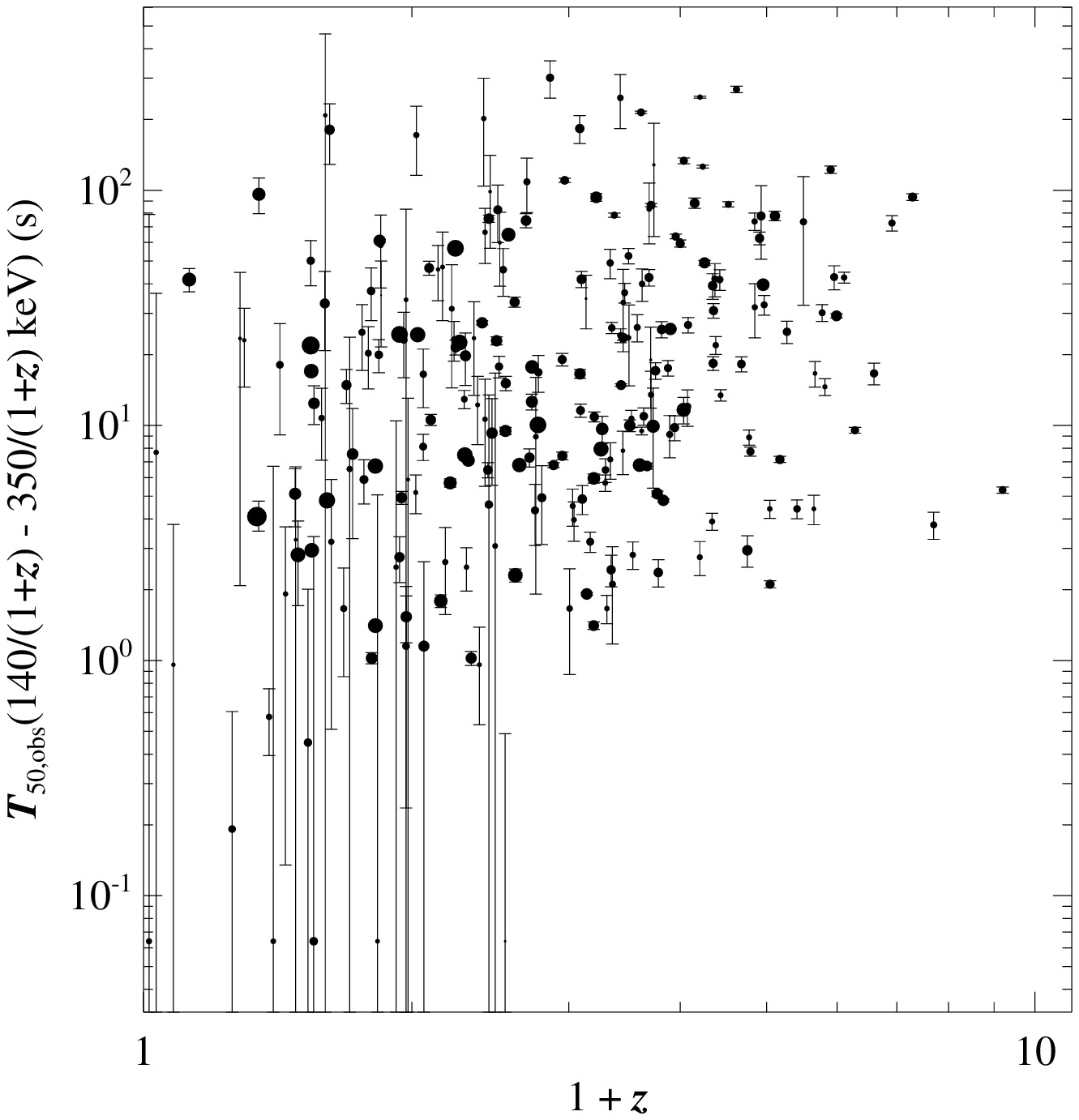}
    \\[-2ex]
    \includegraphics[width=7.5cm,height=7.2cm,angle=0,clip]{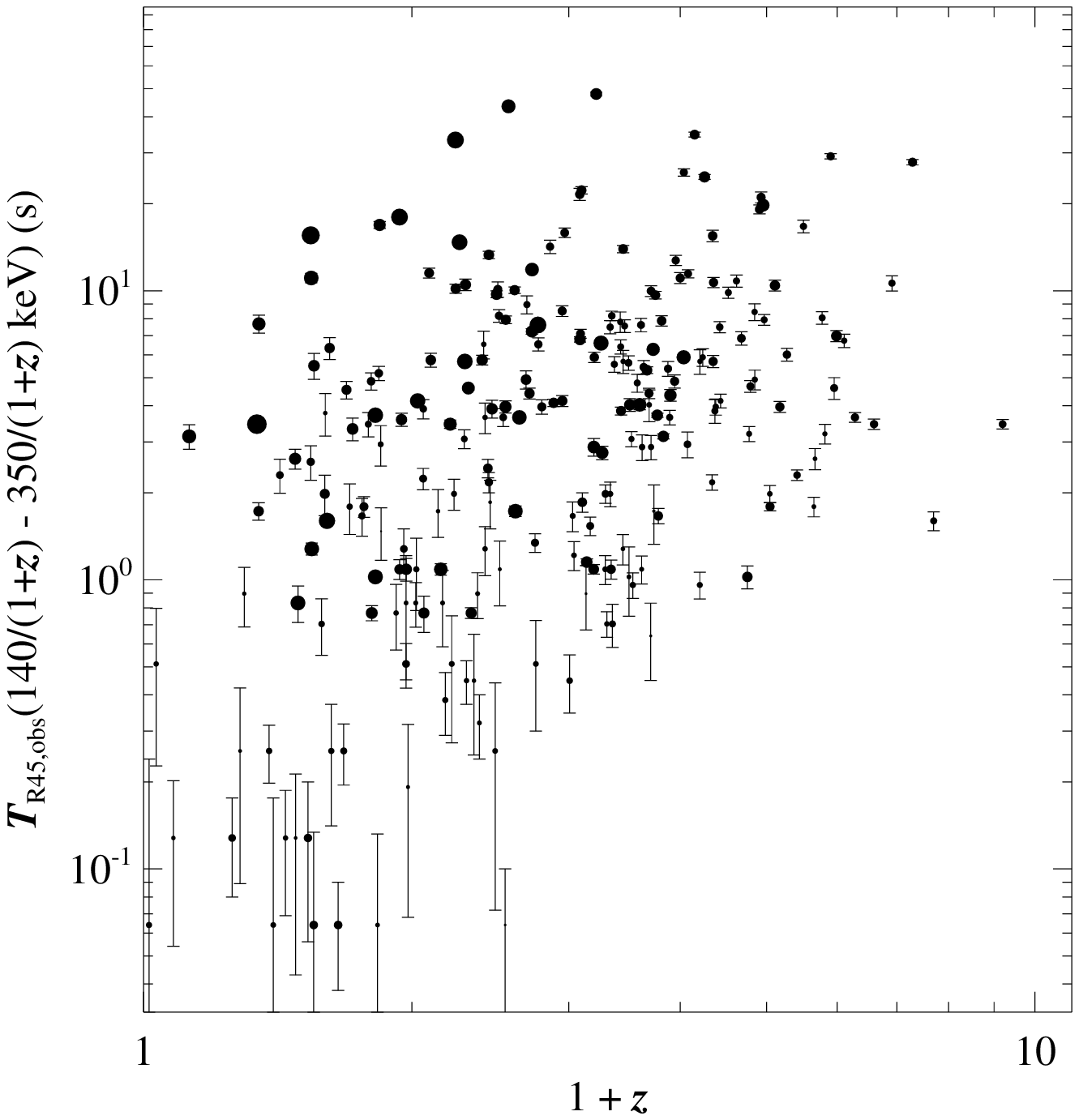}
    \\[-2ex]
  \end{center}
  \caption{Calculated GRB durations in the \citet{2013ApJ...778L..11Z}
    $140/  \left( 1 +  z \right)$--$350  / \left(  1 +  z \right)$~keV
    light curves  as a  function of GRB  redshift. Top  panel: $T_{\rm
      90,obs}$, middle panel:  $T_{\rm 50,obs}$, bottom panel: $T_{\rm
      R45,obs}$.     Point   sizes    are   scaled    by   15--350~keV
    signal-to-noise  ratio,  with a  larger  point  size indicating  a
    higher signal-to-noise ratio.}
  \label{fig:dur_vs_z_snr}
\end{figure}

To estimate the total noise during $T_{\rm 90,obs}$, we took a measure
of the  typical background in  the $15-350$~keV energy range  from the
online repository  detailed in \citet{2007ApJ...671..656B}.   As such,
we determined $B  \approx 8000$~counts.s$^{\rm -1}$, which corresponds
to the count  rate expected to be observed  by \textit{Swift}/BAT from
the          cosmic           X-ray          background          (CXB;
\citealt{2008ApJ...689..666A,1999ApJ...520..124G}).   This  background
rate  was assumed  to  be constant,  converted  to a  value per  fully
illuminated  detector  and integrated  over  the  duration of  $T_{\rm
  90,obs}$.   Figure \ref{fig:dur_vs_z_snr}  shows all  three duration
measures  as  a function  of  redshift.  The  size  of  each point  is
determined according to the calculated signal-to-noise ratio.\par

Immediately evident is that  high-redshift GRBs typically have a lower
signal-to-noise  ratio. This  is  expected, as  GRBs  of an  identical
intrinsic luminosity will appear fainter to \textit{Swift}/BAT when at
a  greater  luminosity  distance.   Another  apparent  trend,  is  the
decreasing significance of burst signal-to-noise ratio with decreasing
duration at any given redshift. That is, at any single redshift, it is
more  difficult to  detect a  relatively shorter  duration  GRB.  This
suggests that  the high-redshift, short-duration region  of the Figure
\ref{fig:dur_vs_z_snr} may be under-populated primarily as a result of
detector sensitivity.   We can also confirm that  the lowest duration,
low redshift GRBs have low signal-to-noise ratios, as expected.\par

\begin{figure}
  \begin{center}
    \includegraphics[width=7.5cm,height=7.2cm,angle=0,clip]{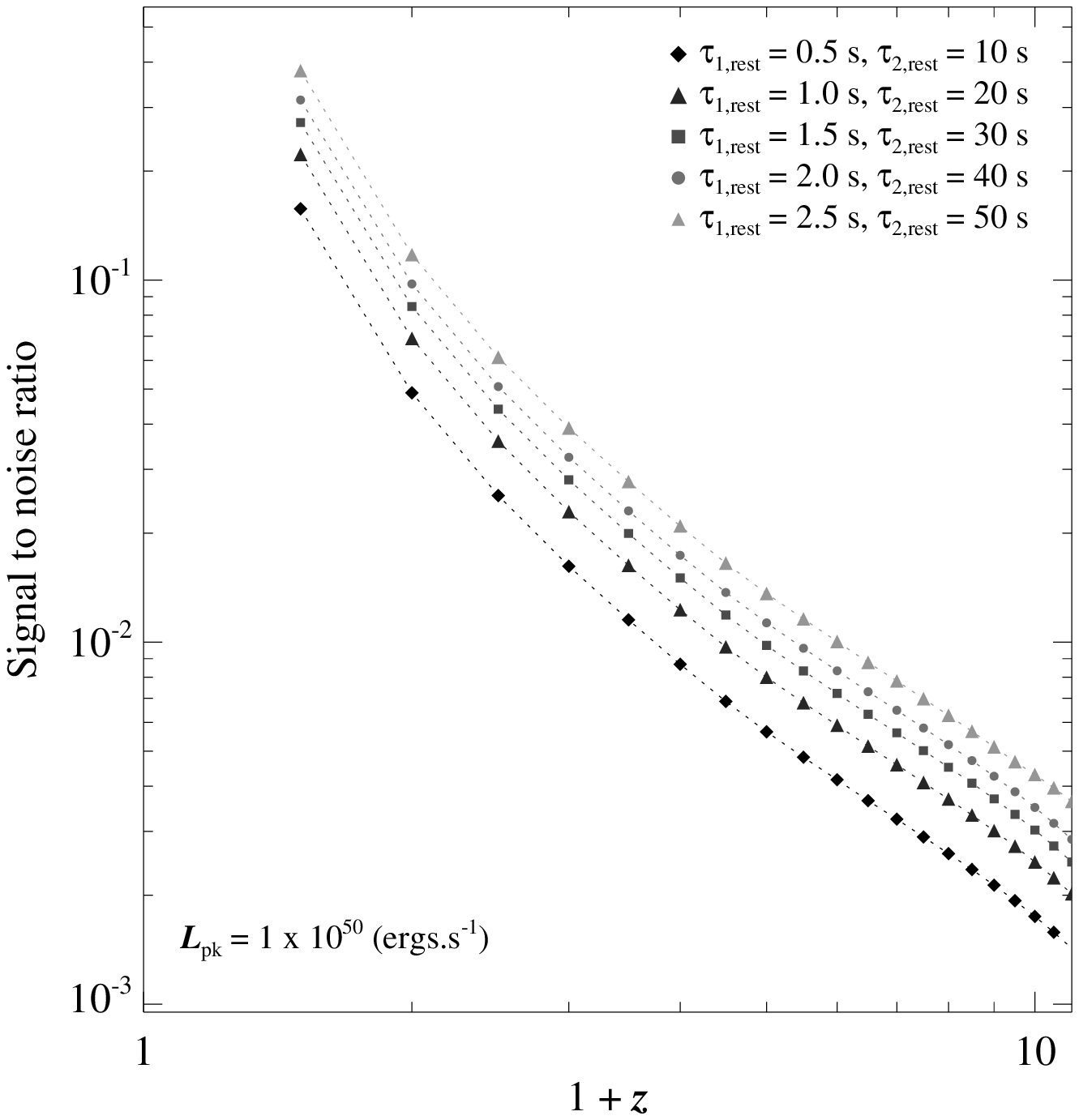}
    \\[-2ex]
    \includegraphics[width=7.5cm,height=7.2cm,angle=0,clip]{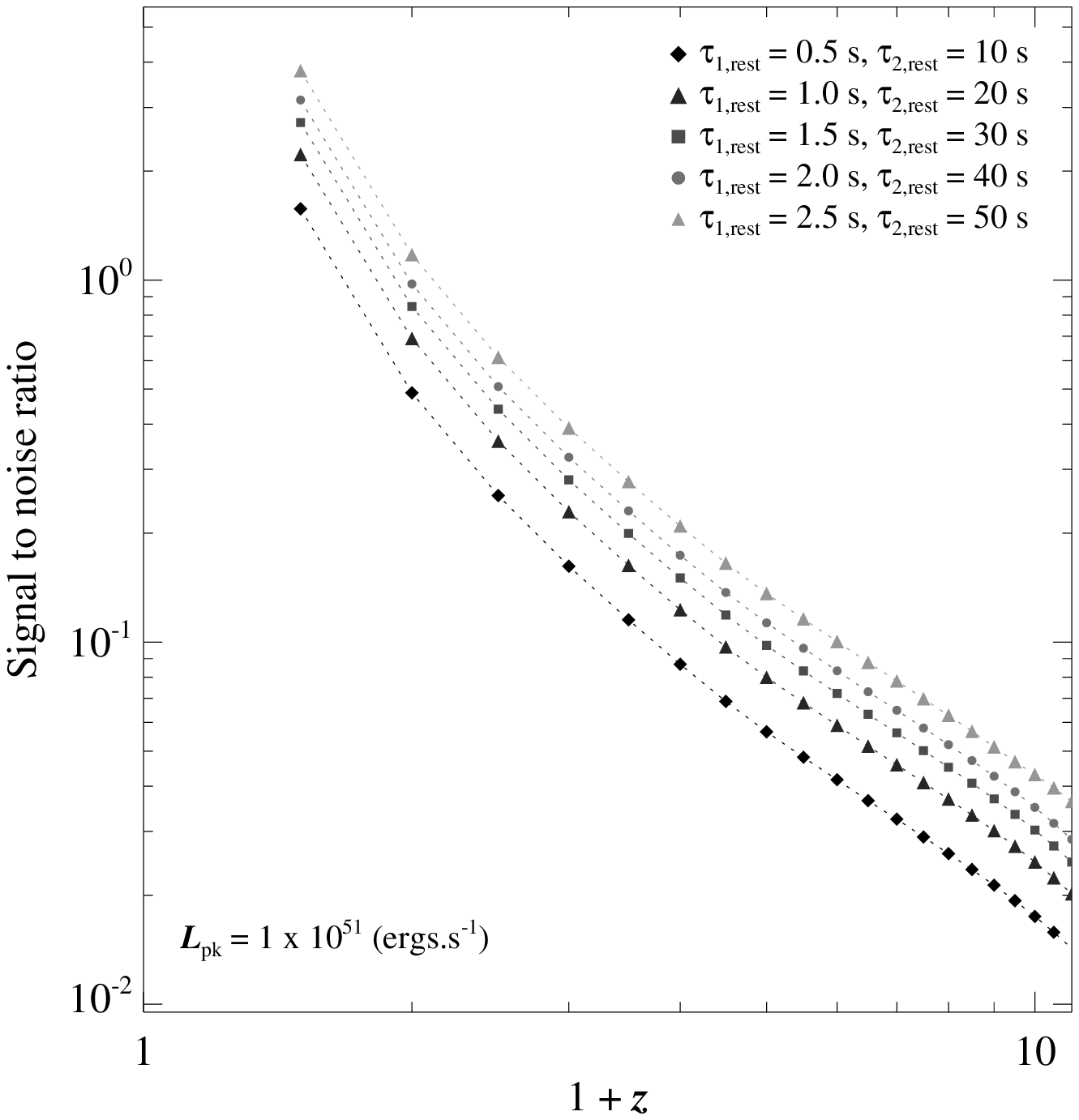}
    \\[-2ex]
    \includegraphics[width=7.5cm,height=7.2cm,angle=0,clip]{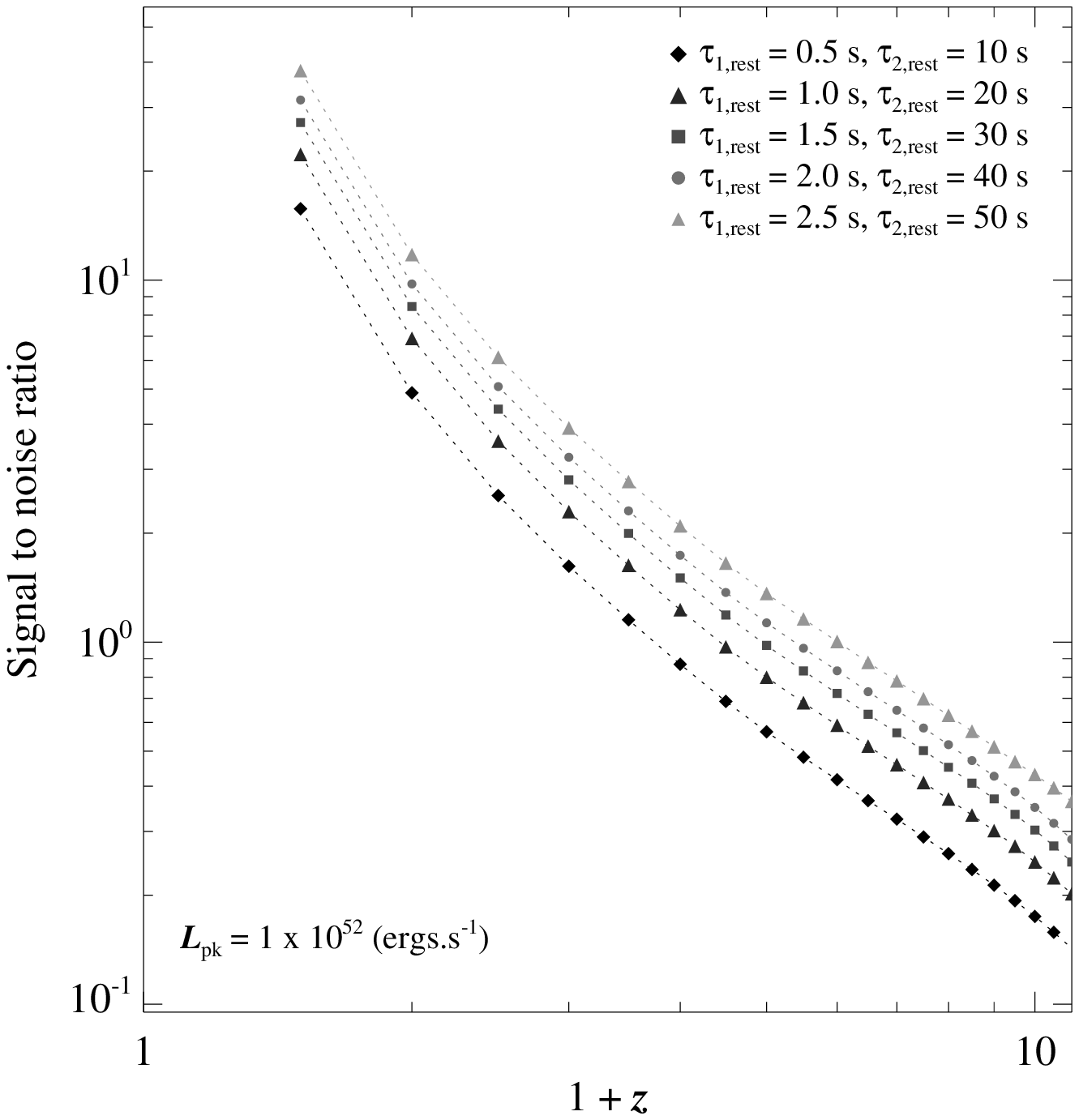}
    \\[-2ex]
  \end{center}
  \caption{Signal   to   noise   ratio    of   $140/\left(   1   +   z
    \right)$--$350/\left( 1 + z \right)$~keV light curves for simulated
    pulses. Top panel: $L_{\rm  pk}$ = $1 \times 10^{50}$ ergs.s$^{\rm
      -1}$,  middle   panel:  $L_{\rm   pk}$  =  $1   \times  10^{51}$
    ergs.s$^{\rm -1}$, $L_{\rm pk}$  = $1 \times 10^{52}$ ergs.s$^{\rm
      -1}$.}
  \label{fig:mod_snr_vs_z}
\end{figure}

To  quantify   the  effects  of  GRB  duration   on  detectability  by
\textit{Swift}/BAT, we calculated the signal-to-noise ratio of several
model  GRBs.   Our model  GRBs  comprised of  a  single  pulse with  a
morphological    shape    described    by    the    prescription    of
\citet{2005ApJ...627..324N}.    In  this   case,  each   pulse   is  a
combination  of  rising   and  declining  exponentials.   We  produced
model-only  $140/\left(1+z\right)$--$350/\left(  1  +  z  \right)$~keV
light curves for  five different duration pulses.  The  details of the
pulse  durations   are  given  in   each  of  the  panels   of  Figure
\ref{fig:mod_snr_vs_z}.  We  then defined  the total duration  of each
model light curve as being when  the pulse profile exceeded 5\% of the
peak flux value.\par

To estimate the correct number  of background counts, we converted the
estimate for the full $15$--$350$~keV light curve to the $140/\left( 1
+ z  \right)$--$350/\left( 1 + z  \right)$~keV range  using the online
Portable          Interactive          Multi-Mission         Simulator
(PIMMS)\footnote{http://heasarc.gsfc.nasa.gov/Tools/w3pimms.html}.\par

We normalised  each pulse, such  that it corresponded to  an intrinsic
peak   luminosity   of   $L_{\rm   pk}$~=~10$^{50}$,   10$^{51}$   and
10$^{52}$~ergs.s$^{-1}$. The results of these simulations are shown in
Figure \ref{fig:mod_snr_vs_z}.\par

Figure  \ref{fig:mod_snr_vs_z}  demonstrates the  two  trends seen  in
Figure  \ref{fig:dur_vs_z_snr}. First,  signal-to-noise  ratio reduces
with increasing  redshift for any  given pulse. More  importantly, the
total  signal-to-noise  ratio at  any  given  redshift increases  with
increasing pulse duration. It is  important to note, however, that the
values of  signal-to-noise ratio do  not directly compare to  those at
which  the  \textit{Swift}/BAT  trigger,   due  to  a  varying  energy
range. Notably, at high redshifts, the width of the light curve energy
band decreases significantly as it is defined in the rest frame of the
GRB.\par

An underlying feature of the observed sample of GRBs available in this
work    is    that    they    must   initially    be    detected    by
\textit{Swift}/BAT.  This  means  that,  despite using  a  rest  frame
defined  energy band  to  mitigate the  energy  dependence within  the
measured  durations,  the  sample  is  fundamentally  defined  in  the
observer frame by the triggering criteria of \textit{Swift}/BAT.\par

Despite \textit{Swift}/BAT having a  plethora of trigger criteria, the
majority of \textit{Swift}/BAT triggered GRBs are the result of ``rate
triggers'', which respond to a rapid increase in flux as quantified by
a signal-to-noise  ratio.  As shown  in Figure \ref{fig:mod_snr_vs_z},
such a  signal to noise  ratio threshold effectively corresponds  to a
threshold   in  duration   for   a  burst   with   a  given   observed
brightness.\par

If we consider that the \textit{Swift}/BAT is intrinsically limited in
the durations  that can be detected  in the observer  frame, such that
only  bursts with  $T_{\rm 90,obs}\left(  15 -  35 {\rm  keV}\right) >
T_{\rm  90,critical}\left(  15  -   35  {\rm  keV}\right)$  result  in
triggered events, we can show how this would translate into a limit in
the duration measured in a rest frame defined energy band.\par

As      $T_{\rm      90,obs}      \propto     E_{\rm      obs}^{-0.4}$
\citep{1996ApJ...459..393N}, we can convert $T_{\rm 90,critical}$ to a
duration  threshold  in a  rest  frame  defined  energy band,  $T_{\rm
  90,critical}^{\prime}$:

\begin{equation}
  T_{\rm 90,critical}^{\prime} = T_{\rm 90,critical} \left( \frac{E^{\prime}}{E_{\rm obs}} \right)^{-0.4},
  \label{eq:crit_1}
\end{equation}

where $E^{\prime}$  and $E_{\rm  obs}$ are characteristic  energies of
the  rest  frame defined  and  observer  frame  defined energy  bands,
respectively. In Equation \ref{eq:crit_1}, both energies have to be in
a common frame of reference,  and so in the observer frame $E^{\prime}
= E_{\rm rest} / \left( 1 + z \right)$. $E_{\rm rest}$ is the standard
characteristic energy of the rest frame defined energy channel, in the
rest  frame of  the GRB.  Substituting this  definition  into Equation
\ref{eq:crit_1} yields:\par

\begin{equation}
  T_{\rm 90,critical}^{\prime} = T_{\rm 90,critical} \left( \frac{E_{\rm rest}}{\left( 1 + z \right) E_{\rm obs}} \right)^{-0.4}.
  \label{eq:crit_2}
\end{equation}

In  Equation \ref{eq:crit_2}  three quantities  are  constant: $T_{\rm
  90,critical}$,  $E_{\rm rest}$  and $E_{\rm  obs}$, thus  it follows
that   $T_{\rm   90,critical}^{\prime}    \propto   \left(   1   +   z
\right)^{0.4}$.   Given a  minimum  value of  duration  defined in  an
observer frame  energy band,  that can result  in a  detectable burst,
there is  a related  limit in  a rest frame  defined energy  band.  As
shown   in  Equation  \ref{eq:crit_2},   this  value   increases  with
increasing redshift of the GRB  as the conversion between the observer
frame defined energy  band and rest frame defined  energy band becomes
large.   Thus  such  an   effect  censors  the  data,  preventing  the
measurement of high redshift,  shorter duration GRBs. By censoring the
parameter  space in  this  way, an  artificial  signal of  a trend  is
introduced,  such  that it  might  appear  that $T_{\rm  90,obs}\left(
E_{\rm 1,rest} / \left( 1 + z  \right) - E_{\rm 2,rest} / \left( 1 + z
\right) \right) \propto \left( 1 + z \right)^{0.4}$.\par

The significance  of any trend of increasing  duration with increasing
redshift above the  null value of $\left( 1 +  z \right)^{0.4}$ can be
estimated using  Equation \ref{eq:trend_sig}, where $I$  is the fitted
index, with a reported uncertainty of  $\Delta I$ and $I_{0} = 0.4$ is
the null value expected as a result of censorship:\par

\begin{equation}
  \sigma = \frac{|I - I_{0}|}{\Delta I}.
  \label{eq:trend_sig}
\end{equation}

Applying Equation \ref{eq:trend_sig} to the fitted indices reported in
Table  \ref{tab:weed_fit_details} we find  that 4  of the  6 relations
have significances less than one sigma. When considering the geometric
average of $T_{\rm  50,obs}\left( 1 + z \right)$  for the full sample,
the putative correlation has a significance of 1.8$\sigma$, while that
of  the  full sample  for  $T_{\rm  R45,obs}$  has a  significance  of
3.4$\sigma$. In all cases, this significance reduces when applying the
brightness  threshold used  by  \citet{2013ApJ...778L..11Z}.  This  is
because  such a  threshold increases  the level  of censorship  of the
data. The indices in all cases move to values compatible with $I_{0} =
0.4$.\par

\subsection{Sampling the brightest GRBs}
\label{sec:bright_subset}

We also considered only the  brightest GRBs in the full sample. Unlike
\citet{2013ApJ...778L..11Z}, we considered the average signal-to-noise
ratio as  defined in \S~\ref{sec:hi_z_short_dur}  as this is  a better
indicator  of the  total brightness  of the  prompt  emission, whereas
$F_{\rm pk}$  is biased towards only  the brightest region  of a light
curve.\par

\begin{figure}
  \begin{center}
    \includegraphics[height=7.2cm,width=7.5cm,angle=0,clip]{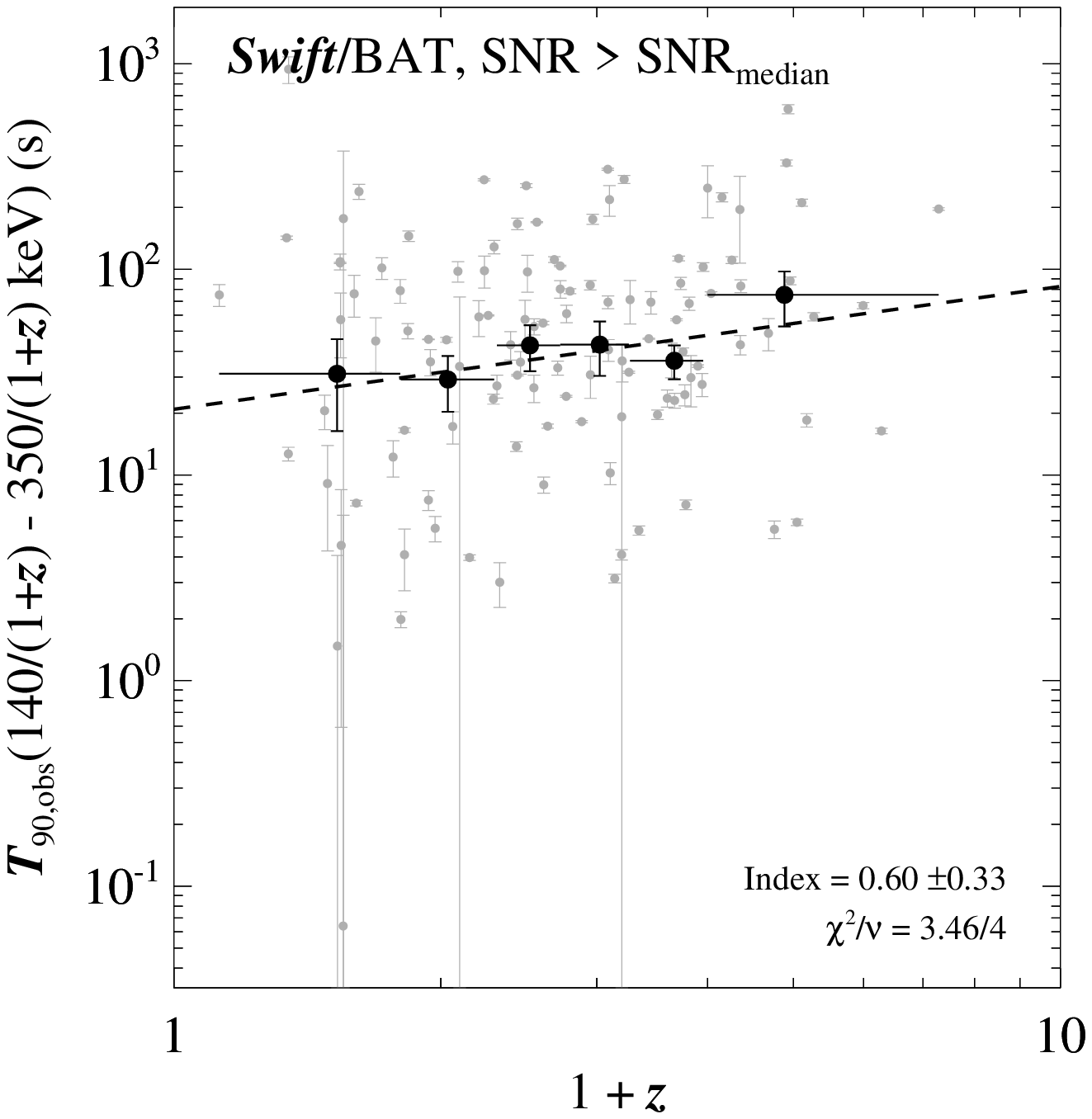}
    \\[-2ex]
    \includegraphics[height=7.2cm,width=7.5cm,angle=0,clip]{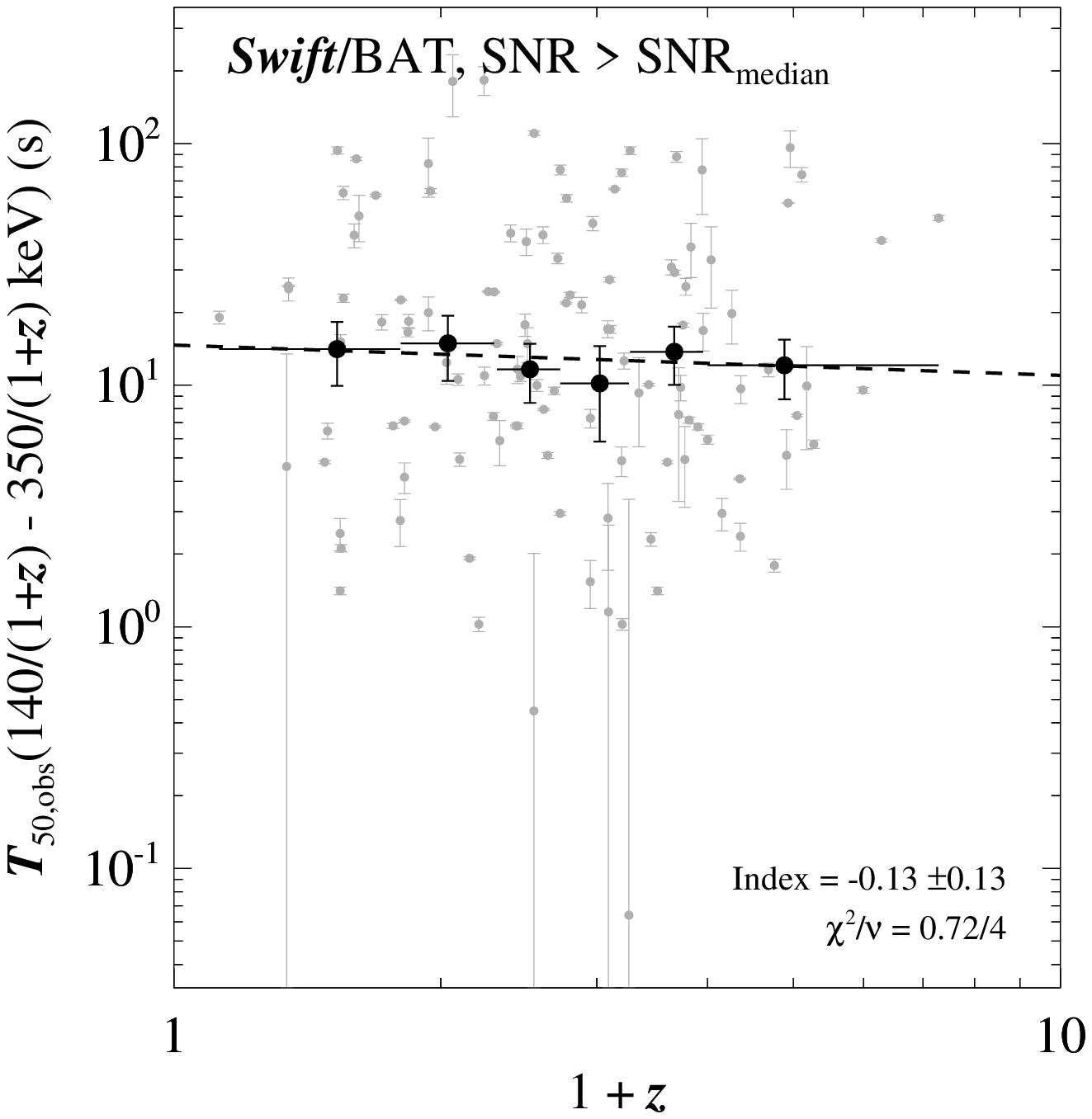}
    \\[-2ex]
    \includegraphics[height=7.2cm,width=7.5cm,angle=0,clip]{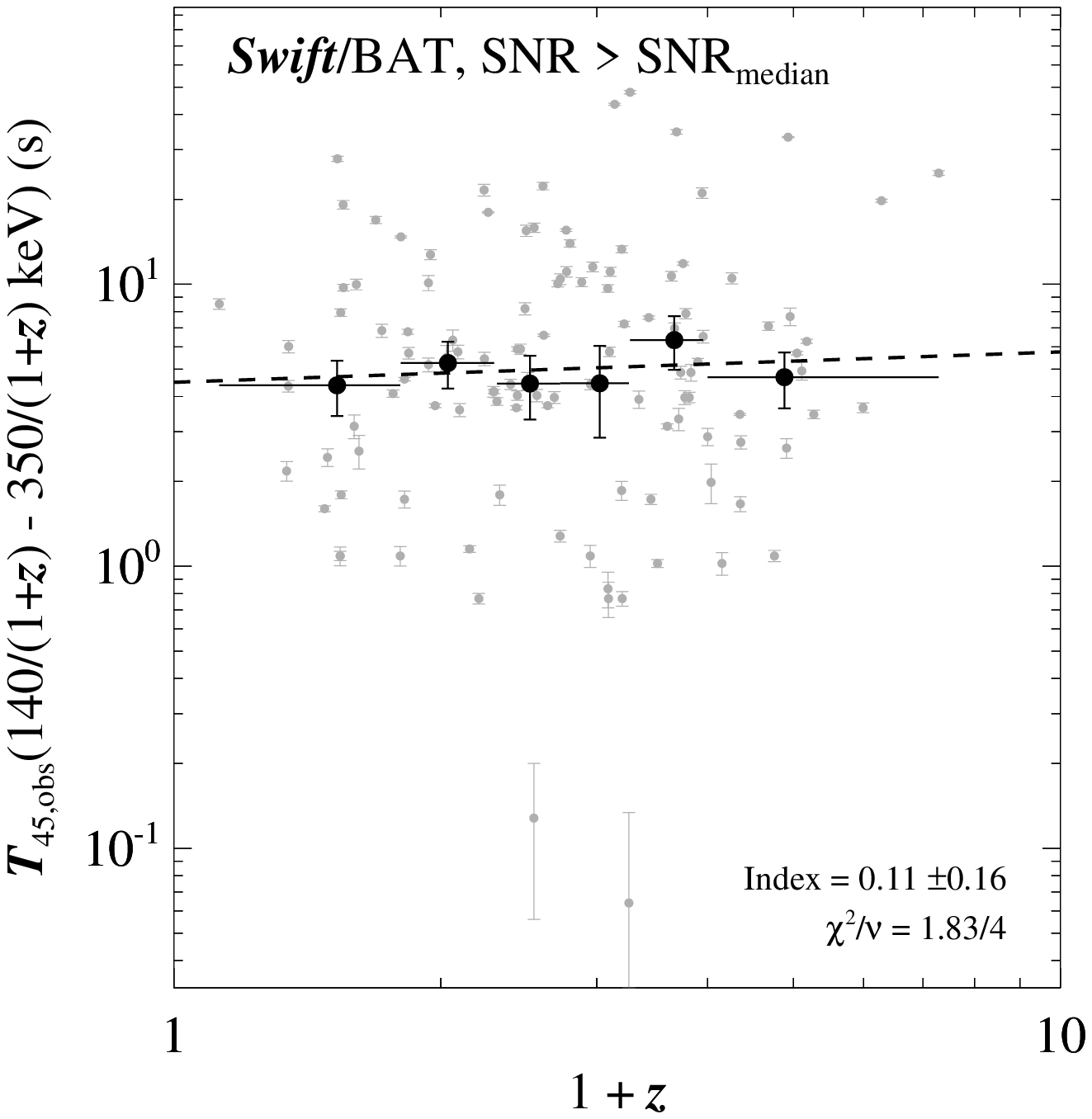}
    \\[-2ex]
  \end{center}
  \caption{$T_{\rm  90,obs}$, $T_{\rm  50,obs}$ and  $T_{\rm R45,obs}$
    obtained  in the  $140/\left(1  + z  \right)$--$350/\left(  1 +  z
    \right)$~keV  energy range for  116 highest  signal-to-noise ratio
    GRBs in  the full  \textit{Swift}/BAT.  Grey points  correspond to
    the individual GRBs, while black points are the geometric average.
    Binning of GRBs is identical in all three panels. The black dashed
    line  in each  panel  corresponds  to a  power-law  fitted to  the
    average bins.   The $\chi^{2}$ fit statistics,  degrees of freedom
    and power-law  indices of all  three models are indicated  in each
    panel.}
  \label{fig:bright_dur_only}
\end{figure}

We took all bursts with a signal-to-noise ratio above the median value
and  a measurable  value for  all three  durations considered  in this
work. We binned the data according  to redshift and performed a fit to
the  geometric  average  of  each  duration  measure  as  before.  The
durations   as   functions  of   redshift   are   plotted  in   Figure
\ref{fig:bright_dur_only}.\par

\begin{table}
  \centering
  \caption{Details  of fits  to  geometric average  of  duration as  a
    function   of    $\left(   1+z\right)$   for    the   bright   GRB
    sample. $N_{\rm GRBs}$ are the  number of bursts contained in all
    six bins, while $\log_{10}N$ is the logarithm of the normalisation
    to each fitted power-law.}
  \label{tab:bright_fit_details}
  \begin{tabular}{ccccc}
    \hline
    \hline
    Duration & $N_{\rm GRBs}$ & $\log_{10}N$ & Index & $\chi^{2} / \nu$ \\
    \hline
    $T_{\rm 90,obs}$ & 116 & 1.32 $\pm$0.17 & 0.60 $\pm$0.33 & 3.46/4 \\
    $T_{\rm 50,obs}$ & 116 & 1.17 $\pm$0.06 & -0.13 $\pm$0.13 & 0.72/4 \\
    $T_{\rm RT45,obs}$ & 116 & 0.65 $\pm$0.07 & 0.11 $\pm$0.16 & 1.83/4 \\
    \hline
  \end{tabular}
\end{table}

As  shown  in  Table  \ref{tab:bright_fit_details}, the  evolution  of
$T_{\rm  90,obs}\left( {\rm  SNR}  > {\rm  SNR}_{\rm median}  \right)$
reproduces  a very  similar best  fit model  to the  bright  sample as
defined  by  $F_{\rm  pk}$  outlined in  \S~\ref{sec:zhang_ext}.   The
power-law  index  retrieved  from  fitting the  geometric  average  to
$T_{\rm 90,obs}\left(  {\rm SNR} > {\rm SNR}_{\rm  median} \right)$ is
0.60~$\pm$0.33,  giving  a significance  of  only  0.6$\sigma$ to  the
trend.\par

However, the fits obtained for  $T_{\rm 50,obs}\left( {\rm SNR} > {\rm
  SNR}_{\rm median}  \right)$ and  $T_{\rm R45,obs}\left( {\rm  SNR} >
{\rm  SNR}_{\rm median}  \right)$ are  consistent, within  error, with
being constant with redshift, which would imply that $T_{\rm 50,rest}$
and  $T_{\rm R45,rest}$  in the  brightest half  of \textit{Swift}/BAT
GRBs reduce with increasing redshift, such that $T_{\rm 50,rest}\left(
{\rm  SNR} >  {\rm  SNR}_{\rm median}  \right)  \propto \left(  1 +  z
\right)^{-1}$. Considering the errors in the fitted indices, which are
of   the  same   order  of   magnitude  as   the   indices  themselves
(-0.13~$\pm$0.13 for $T_{\rm  50,obs}$, and 0.11~$\pm$0.16 for $T_{\rm
  R45,obs}$), this  is not  likely a physical  effect, as  it suggests
correlations strengths of $\lesssim$1$\sigma$.\par

\section{Conclusions}
\label{sec:conc}

In  this  work  we  have  investigated whether  duration  measures  of
\textit{Swift}/BAT  and \textit{Fermi}/GBM  detected GRBs  exhibit the
effects of cosmological time dilation.  We first verify the results of
\citet{2013ApJ...778L..11Z} and investigate  which method of averaging
individual  durations   is  the  most   robust  as  shown   in  Figure
\ref{fig:zhang_all}. As  a power-law model is employed,  we choose the
geometric average.\par

We  find that,  when accounting  for  the measured  errors in  $T_{\rm
  90,obs}$,  a  power-law  is  a  statistically  unacceptable  fit  to
individual bright GRBs  as shown in Figure \ref{fig:bright_zhang_ind}.
We then updated the \textit{Swift}/BAT sample to include an additional
93   GRBs  that  have   occurred  since   the  original   analysis  by
\citet{2013ApJ...778L..11Z}.  Using this total sample of 232 bursts we
investigate  the  evolution  of  average durations  $T_{\rm  90,obs}$,
$T_{\rm 50,obs}$  and $T_{\rm R45,obs}$  as a function of  redshift in
the   $140/\left(1+z\right)$--$350/\left(1+z\right)$~keV  range.   All
three  durations  exhibit  a   trend  of  increasing  with  increasing
redshift.  The power-law  indices of  these trends  were  reduced when
filtering out bursts with poorly measured values of duration.\par

The model  power-law index obtained  fitting the geometric  average of
$T_{\rm 50,obs}\left( 140/\left(1+z\right) - 350/\left(1+z\right)~{\rm
  keV}\right)$  has a value  most consistent  with that  expected from
time-dilation.  We  do find,  however, that the  large scatter  of the
distribution of individual duration  values is large, leading to large
statistical errors on each average bin, and therefore a large error in
the fitted parameters of the power-law model.\par

We also compare the distributions  of all three durations, in the rest
frame, both above and below  the median redshift $z_{\rm median}$, and
within the upper and lower  quartiles of redshift. Using the Student's
$t$-test we  find that the  distributions of durations  are consistent
with having the same mean value in five out of six cases. We also find
a          3$\sigma$           difference          in          $T_{\rm
  R45,obs}\left(140/\left(1+z\right)-350/\left(1+z\right)~{\rm
  keV}\right)/\left(1+z\right)$ distributions  above and below $z_{\rm
  median}$.  Of  the three duration  measures, it is  perhaps expected
that $T_{\rm  R45,obs}$ would show the least  evidence of cosmological
time dilation.  $T_{\rm R45,obs}$  contains only the brightest regions
of a light  curve. Conversely, should any quiescent  period be present
between  prompt pulses,  $T_{\rm  90,obs}$ and  $T_{\rm 50,obs}$  will
contain  this  time.   \citet{2013ApJ...765..116K} propose  that  such
quiescent periods between pulses are  likely to be the best tracers of
cosmological time dilation.\par

We   cross-referenced    the   232   \textit{Swift}/BAT    GRBs   with
\textit{Fermi}/GBM  data, to find  that 57  bursts with  redshift have
been  detected  with   both  instruments.   Figure  \ref{fig:all_durs}
demonstrates that this sample is dominated by small number statistics.
As such the \textit{Swift}/BAT data  in this subset of bursts does not
reflect the trends  shown for the full sample.  For all three duration
measures, the  obtained \textit{Fermi}/GBM seem to  evolve more weakly
with  redshift  than  those  obtained  for  the  same  GRBs  with  the
\textit{Swift}/BAT. The  significance of this difference  is not high,
however, once more due to the small size of the joint sample.\par

We finally consider  the origin of the apparent  durations trends as a
function of redshift,  to assess whether the physical  origins are due
to the cosmological time dilation  of a common rest frame distribution
of GRBs.  We find no evidence  of under-sampling of  the long duration,
low redshift region of the parameter space.\par

We then demonstrate the dearth of high-redshift, short duration (where
short refers to the low duration end of the long GRB distribution) can
be attributed  to censorship of the parameter  space.  This censorship
arises  from originally  creating the  sample of  detected GRBs  in an
observer frame energy band, as this is how \textit{Swift}/BAT triggers
are defined.   As BAT triggers have a  signal-to-noise ratio threshold
and, for a given peak flux, shorter GRBs have lower significance, this
naturally places a minimum limit of detectable duration for a long GRB
of a given  brightness.  This limit can be  converted to an equivalent
limit in  the duration distribution  of the rest frame  defined energy
band.  As the  difference between the observer frame  defined and rest
frame  defined energy  bands increases  with increasing  redshift, the
lower limit in detectable  durations rises accordingly.  Thus the high
redshift,  low duration  region of  the parameter  space  suffers from
censorship,  which  helps  to   artificially  induce  a  signature  of
cosmological  time  dilation  in  the duration-redshift  plane.   This
censorship increases  as $\left( 1  + z \right)^{0.4}$, thus  giving a
null hypothesis value for the power-law index of any duration-redshift
relation.  Assessing  the significance of the relations  found in this
work, we find that they typically are less than 1 or 2$\sigma$.\par

Finally,   we  isolate   bursts  that   are  well   detected   by  the
\textit{Swift}/BAT instrument.  Our  metric for ``brightness'' is also
an average signal-to-noise ratio  over the entire burst duration, thus
ensuring the pulse tail is well sampled. In so doing, we find that the
geometric  average  of  $T_{\rm  90,obs}$  may  still  correlate  with
redshift,  although  $T_{\rm 50,obs}$  and  $T_{\rm  R45,obs}$ do  not
appear to.  The  reason for the difference between  the three measures
may result  from $T_{\rm 90}$ being defined  in such a way  that it is
more likely  to include  any quiescent periods  between pulses  in the
prompt light curve.   These will very strongly exhibit  the effects of
cosmological  time-dilation. Care  must  be taken  when imposing  such
thresholds,  however,  as  they  potentially  enhance  the  censorship
effects  previously  discussed,  by  strengthening the  thresholds  in
signal-to-noise ratio  any GRB  must satisfy to  be considered  in the
sample.\par

This work highlights the relative merits of each of the three duration
measures. Of  the three, $T_{\rm  90}$ comes the closest  to capturing
the total  prompt duration. However, in  doing so, it  is necessary to
deeply probe the tail of pulse emission. The uncertainty, therefore in
determining the end of the  $T_{\rm 90}$ duration can be high. $T_{\rm
  R45}$ only  captures a  sense of the  brightest regions of  a burst.
Initially, this seems a promising prospect for extracting cosmological
time dilation  if one  assumes pulses to  be self-similar in  the rest
frame.  However,  the population of pulses within  prompt light curves
is  diverse \citep{2005ApJ...627..324N}.  Additionally,  $T_{\rm R45}$
does  not  contain  information  concerning the  frequency  of  bright
emission periods. That  is to say, without further  information, it is
not  known if  $T_{\rm R45}$  is continuous,  or comprised  of several
shorter episodes.\par

$T_{\rm 50}$ could  be considered to offer more  information than both
$T_{\rm  90}$ and  $T_{\rm R45}$  can individually.  By  considering a
narrower  region of  the cumulative  distribution of  prompt emission,
$T_{\rm 50}$ samples the pulse tail  to a point that is better defined
in  signal-to-noise  ratio. However,  $T_{\rm  50}$  still provides  a
better  estimate of  total prompt  duration when  compared  to $T_{\rm
  R45}$ as it can include intervening quiescent periods between bright
pulses.\par

Given  the  scatter  of   the  distributions  of  durations,  and  the
unavoidable censorship  of the redshift-duration  parameter space, the
quest for a redshift or luminosity indicator within high-energy prompt
GRB light curves seems unlikely to yield a positive result.\par

\begin{bibliography}{ms}
  \bibliographystyle{mn2e}
\end{bibliography}

\label{lastpage}
\end{document}

%% file: aas_abbrev.tex
\newcommand\apjl{ApJ}%
\newcommand\apjs{ApJS}%
\newcommand\ao{Appl.~Opt.}%
\newcommand\aap{A\&A}%